\documentclass[lettersize,journal]{IEEEtran}

\usepackage{multirow}
\usepackage{booktabs}
\usepackage{amsmath,amsfonts,amssymb}
\usepackage{bm}
\usepackage{array}
\usepackage{threeparttable}
\usepackage{algorithmic}
\usepackage{algorithm}
\usepackage{textcomp}
\usepackage{stfloats}
\usepackage{verbatim}
\usepackage{graphicx}
\usepackage{subcaption}
\usepackage{xcolor}
\usepackage{cite}
\usepackage{url}
\usepackage{booktabs}
\usepackage{multirow}
\usepackage{microtype}
\usepackage{balance}
\usepackage[colorlinks=true,citecolor=blue,linkcolor=blue,urlcolor=blue]{hyperref}
\usepackage{orcidlink}
\usepackage{tikz}
\usepackage{multirow}
\usepackage{pifont}
\usepackage{graphicx}

\newcommand{\cmark}{\ding{51}}
\newcommand{\xmark}{\ding{55}}
\usetikzlibrary{arrows.meta,positioning,fit}

\graphicspath{{figures/}}
\definecolor{todored}{RGB}{170,25,35}

\newcommand{\vect}[1]{\boldsymbol{#1}}
\newcommand{\mat}[1]{\mathbf{#1}}

\title{Manifold-Constrained Residual Adaptation with Frame-Adaptive Layer Aggregation for Speaker Verification}

\author{Zezhong~Jin\orcidlink{0009-0001-0828-2933},
Xiaoyu~Wang\orcidlink{0009-0008-7612-2051},
Zhe~Li\orcidlink{0000-0002-0519-7434},
Chong-Xin~Gan\orcidlink{0000-0003-2121-5935},
Siqing~Qin\orcidlink{0009-0002-0786-4859},
Man-Wai~Mak\orcidlink{0000-0001-8854-3760},~\IEEEmembership{Senior Member,~IEEE},
and Kong~Aik~Lee\orcidlink{0000-0001-9133-3000},~\IEEEmembership{Senior Member,~IEEE}%
\IEEEcompsocitemizethanks{
\IEEEcompsocthanksitem Z. Jin and X. Wang contributed equally. M.-W. Mak and
K. A. Lee are the corresponding authors. This work was supported in part by
the Research Grants Council of the Hong Kong SAR under Grant 15228223 and by
The Hong Kong Polytechnic University under Project P0049192.
\IEEEcompsocthanksitem Z. Jin, C.-X. Gan, S. Qin, M.-W. Mak, and K. A. Lee
are with the Department of Electronic and Information Engineering, The Hong
Kong Polytechnic University, Hong Kong SAR (e-mail:
zezhong.jin@connect.polyu.hk; man.wai.mak@polyu.edu.hk;
kong-aik.lee@polyu.edu.hk).
\IEEEcompsocthanksitem X. Wang is with Baidu Inc. Z. Li is with the Department of Electrical Engineering, Stanford University, Stanford, CA 94305, USA (e-mail: zhelee@stanford.edu).
}}

\begin{document}
\maketitle

\begin{abstract}
Large self-supervised speech models provide powerful representations for speaker verification, but full fine-tuning is expensive. Most conventional adapters insert small trainable modules to modify the output of each layer, while keeping the original residual connections unchanged. Building on our preliminary study of manifold-constrained Hyper-Connections in speaker encoders, we explore parameter-efficient adaptation of both residual paths and feature selection across layers. We first propose a Manifold-Constrained Residual Adapter (MCRA), which splits the hidden dimensions of each Transformer residual state into $N$ groups and uses a small Sinkhorn-normalized matrix to exchange information among them. A learnable gate controls the balance between the original and mixed residual paths. The gate initially favors the original path, allowing the model to adapt gradually with only one additional parameter at each residual connection. We further propose Frame-Adaptive Layer Aggregation (FALA), an attention-based aggregation method that uses a single learned query shared across all utterances to compute layer weights separately for each speech frame. Because the layer representations depend on the input, the resulting weights vary across both utterances and frames, unlike the fixed weights of a conventional layer-weighted sum. Compared with a frozen WavLM baseline using a fixed layer-weighted sum, our method achieves relative EER reductions of 22.4\% on VoxCeleb1-O and 2.9\% on CN-Celeb, while training only 0.0009\% and 0.0016\% of the WavLM-Large parameters, respectively.
\end{abstract}


\section{Introduction}
\IEEEPARstart{A}{utomatic} speaker verification (ASV) determines whether a
test utterance was produced by a claimed speaker. Modern systems map
variable-length enrollment and test utterances to fixed-dimensional speaker
embeddings and compare them to provide a similarity score \cite{wang2024overview}. Their performance has
advanced through improvements in network architecture, training objective,
and temporal aggregation. Speaker encoders have evolved from x-vector TDNNs
to ResNet- and Res2Net-based models, ECAPA-TDNN, and
MFA-Conformer~\cite{snyder2018xvector,zhou2021resnext_res2net,
desplanques2020ecapa,zhang2022mfaconformer}. Margin-based and metric-learning
objectives improve the separation between speakers~\cite{deng2019arcface,
chung2020metric}. Pooling methods have also progressed from statistics
pooling to attentive and multi-head aggregation~\cite{okabe2018asp,
peng2023mhfa}. These developments have produced increasingly discriminative
speaker embeddings from supervised training data, including phonetic-aware far-field verification by Jin et al.~\cite{jin2024joseph}, attentive spectral aggregation by Tu and Mak~\cite{tu2022attentive_stsp}, and within-global-view knowledge transfer by Jin et al.~\cite{jin2024wgvkt}.

Self-supervised learning (SSL) provides another source of performance gains \cite{jin2024memo,jin2026uamemo}. Models such as wav2vec~2.0, HuBERT, and WavLM are pretrained on large unlabeled speech corpora to learn representations that can be transferred to downstream tasks~\cite{baevski2020wav2vec,
hsu2021hubert,chen2022wavlm}. Their hidden representations can be transferred
to speaker verification by fine-tuning the SSL encoder or by freezing it and
training a speaker-specific backend. Full fine-tuning offers broad adaptation
capacity, but updating hundreds of millions of parameters is computationally demanding. Freezing the
encoder reduces the training cost, but limits task-specific adaptation.

Parameter-efficient fine-tuning (PEFT) seeks a balance between these two
settings. Bottleneck adapters insert small trainable transformations into the
pretrained encoder~\cite{peng2023adapter}. LoRA represents weight updates in
a low-rank subspace~\cite{hu2022lora}. Speaker prompt tuning introduces
trainable tokens, while UniPET-SPK combines prompts and adapters through
learnable gates~\cite{li2024dualpeft,sang2025unipet}. A recent unified study
has further analyzed several PEFT methods for speaker
verification~\cite{li2026unifiedpeft}. Although these methods use different
parameterizations, they mainly adapt layer transformations, model weights, or
task-specific inputs. The residual paths that propagate pretrained features
through the Transformer usually remain unchanged.

Our preliminary work reported in \cite{jin2026speaker_mhc} investigated the residual-path problem
from the perspective of manifold-constrained Hyper-Connections
(mHC). It replaced the residual shortcuts of ResNet-34, Res2Net, and ECAPA-TDNN encoders with multiple
full-width residual streams. A Sinkhorn-normalized, doubly stochastic matrix is used to control information exchange among the streams~\cite{xie2026mhc}. The results established
the effectiveness of constrained residual mixing across several speaker
encoders. However, this study considered only models trained end to end. It
did not address whether residual mixing can be used for PEFT of a frozen SSL
encoder. The original mHC is also difficult to apply directly because it
maintains multiple full-width residual streams, which expands the hidden
state. In addition, PEFT should preserve the pretrained model at the start of
adaptation.

To address this gap, we reformulate constrained residual mixing as a
Manifold-Constrained Residual Adapter (MCRA). MCRA divides the existing
hidden dimension into $N$ feature groups instead of creating $N$ full-width
streams. An $N\times N$ Sinkhorn-normalized matrix mixes information among
the groups. Its parameter count depends on $N$, rather than on the hidden
dimension. A learnable gate interpolates between the original and mixed
residuals. The gate initially favors the original path and gradually
introduces the adapted path during training. MCRA can thus modify residual
information propagation while retaining the frozen backbone and its original
hidden width.

Another challenge is how to combine the intermediate layers of an SSL model.
The conventional layer-weighted sum learns one scalar per layer and
uses the same layer weights for every utterance and speech frame~\cite{yang2021superb}. MHFA
adds input-dependent temporal attention, but its layer weights are still
fixed across utterances and frames~\cite{peng2023mhfa}. LAP computes dynamic
layer weights for each frame, but requires feature projections, multiple
attention heads, and a specialized pooling network~\cite{kim2025lap}. We propose
Frame-Adaptive Layer Aggregation (FALA) as a lightweight alternative. FALA
uses one learned query to compute a different layer distribution for each
speech frame. It requires no query, key, value, or output projection matrix.

This paper substantially extends our preliminary mHC
work~\cite{jin2026speaker_mhc}. The main contributions are summarized as
follows:

\begin{itemize}
    \item We extend manifold-constrained residual mixing from end-to-end
    training of speaker encoders to parameter-efficient
    fine-tuning of frozen SSL models. This study explores residual
    information propagation as a new adaptation target for SSL-based speaker
    verification.
    \item We propose MCRA, which divides the original hidden state into
    feature groups and mixes information among them without increasing the
    hidden dimension. A learnable
    gate preserves the original residual path at initialization and controls
    the contribution of the mixed path during adaptation.
    \item We propose FALA, which assigns frame-dependent weights to the SSL
    layers using a single task-specific query. It replaces the fixed
    layer-weighted sum and adds only a small number of parameters to WavLM-Large.
    \item Extensive experiments with different pretrained models and evaluation
    datasets show that the proposed methods improve speaker verification
    performance while introducing only a few thousand trainable parameters.
\end{itemize}

\section{Related Work}
\subsection{Self-Supervised Learning for Speech and Parameter-Efficient Fine-Tuning}
Speech encoders trained with self-supervised learning (SSL), such as
wav2vec~2.0~\cite{baevski2020wav2vec}, HuBERT~\cite{hsu2021hubert}, and
WavLM~\cite{chen2022wavlm}, provide representations that can be transferred to downstream speaker verification task. The models use different training objectives. wav2vec~2.0 uses a contrastive
objective: for each masked speech segment, it learns to distinguish the
correct speech representation from several alternatives. HuBERT predicts automatically assigned labels for masked speech frames. WavLM combines masked-frame
prediction with denoising, learning to recover speech information from noisy
input. Their hidden layers capture
complementary information at different levels of abstraction, making both
encoder adaptation and layer selection important for downstream speaker
systems. A common approach either fine-tunes the complete SSL encoder or
keeps it frozen and trains a speaker backend such as
ECAPA-TDNN~\cite{desplanques2020ecapa}. The former provides greater
adaptation flexibility but updates hundreds of millions of parameters,
whereas the latter is more economical but leaves the pretrained
representations unchanged.

Parameter-efficient fine-tuning (PEFT) seeks a compromise by freezing most of the
pretrained encoder and optimizing a small task-specific parameter set.
Bottleneck adapters insert compact feature transformations into or alongside
Transformer layers~\cite{peng2023adapter}; LoRA represents updates to
selected weight matrices in a low-rank subspace~\cite{hu2022lora}; and
speaker prompt tuning introduces trainable tokens that guide the pretrained
encoder~\cite{li2024dualpeft}. UniPET-SPK further combines inner- and
inter-layer adapters with deep speaker prompts through learnable
gates~\cite{sang2025unipet}. Recent controlled studies have also compared
these PEFT families under unified speaker-verification settings
~\cite{li2026unifiedpeft}. Despite their different parameterizations, these
methods primarily adapt network weights, intermediate features, or
task-specific inputs while retaining the original residual topology. MCRA
instead learns how information propagates through the residual connections.

\subsection{Hyper-Connections and Manifold-Constrained Residual Routing}
Residual connections use a fixed identity shortcut to propagate information
through deep networks~\cite{he2016resnet,vaswani2017attention}. This
identity path supports optimization and preserves information over depth,
but its connectivity is fixed. Hyper-Connections (HC) extend the identity shortcut
with multiple residual streams and learn how information is combined and
propagated across layers~\cite{zhu2024hyperconnections}. This provides more
flexible connections between successive layers.

When unconstrained HC mixing matrices are composed across many layers, they
may cause signal explosion or vanishing and lead to training instability.
Manifold-constrained Hyper-Connections (mHC) use Sinkhorn normalization to
make the mixing matrix doubly stochastic~\cite{xie2026mhc}, such that every
row and column sums to one. This allows information to be exchanged among
streams while keeping the feature scale stable. Unlike adapters and LoRA,
which modify feature transformations or model weights, HC and mHC change how
features propagate through residual connections.

Our preliminary work was the first to explore mHC for speaker representation
learning. It replaced the residual shortcuts of fully trained Fbank speaker
encoders with static manifold-constrained mixing
~\cite{jin2026speaker_mhc}. Building on this study, the present work explores
whether manifold-constrained residual mixing can be reformulated as a PEFT
method for frozen SSL speech encoders. 

\subsection{Cross-Layer Aggregation for SSL Speaker Verification}
Speaker encoders fine-tuned from pretrained SSL speech models commonly
combine representations from different layers using a learned scalar
weighted sum~\cite{yang2021superb}. This approach learns a global preference
over the layers of the SSL speech model, but it applies
the same layer weights to every utterance and frame. It therefore cannot
adapt layer selection to changing phonetic content or acoustic conditions.

Multi-Head Factorized Attentive Pooling (MHFA) learns separate global layer
weights to form key and value sequences, and then applies multi-head
attention to aggregate information over time~\cite{peng2023mhfa}. Its
temporal attention depends on the input, but its layer weights remain fixed
across utterances and frames. Layer Attentive Pooling (LAP) instead computes
frame-dependent layer weights using feature projections and multiple
attention heads~\cite{kim2025lap}. It enables dynamic layer selection, but
also introduces additional projections, nonlinear operations, and a
specialized pooling backend.

These limitations motivate FALA, a lightweight method that uses one shared
query to assign input-dependent weights to the SSL layers at each frame.

\begin{figure*}[t]
  \centering
  
  \begin{subfigure}[b]{0.24\textwidth} 
    \centering
    \includegraphics[height=5.0cm, width=!]{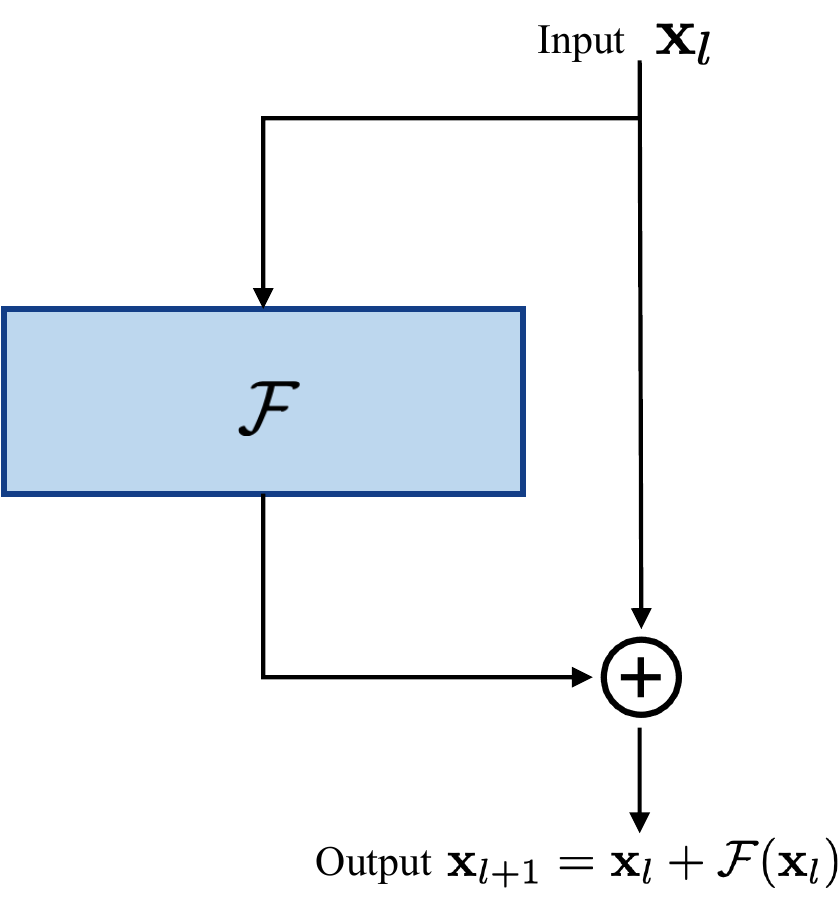} 
    \caption{Standard Residual Connection}
    \label{fig:residual}
  \end{subfigure}
  \hspace{0.001\textwidth}
  \begin{subfigure}[b]{0.31\textwidth}
    \centering
    \includegraphics[height=5.0cm, width=!]{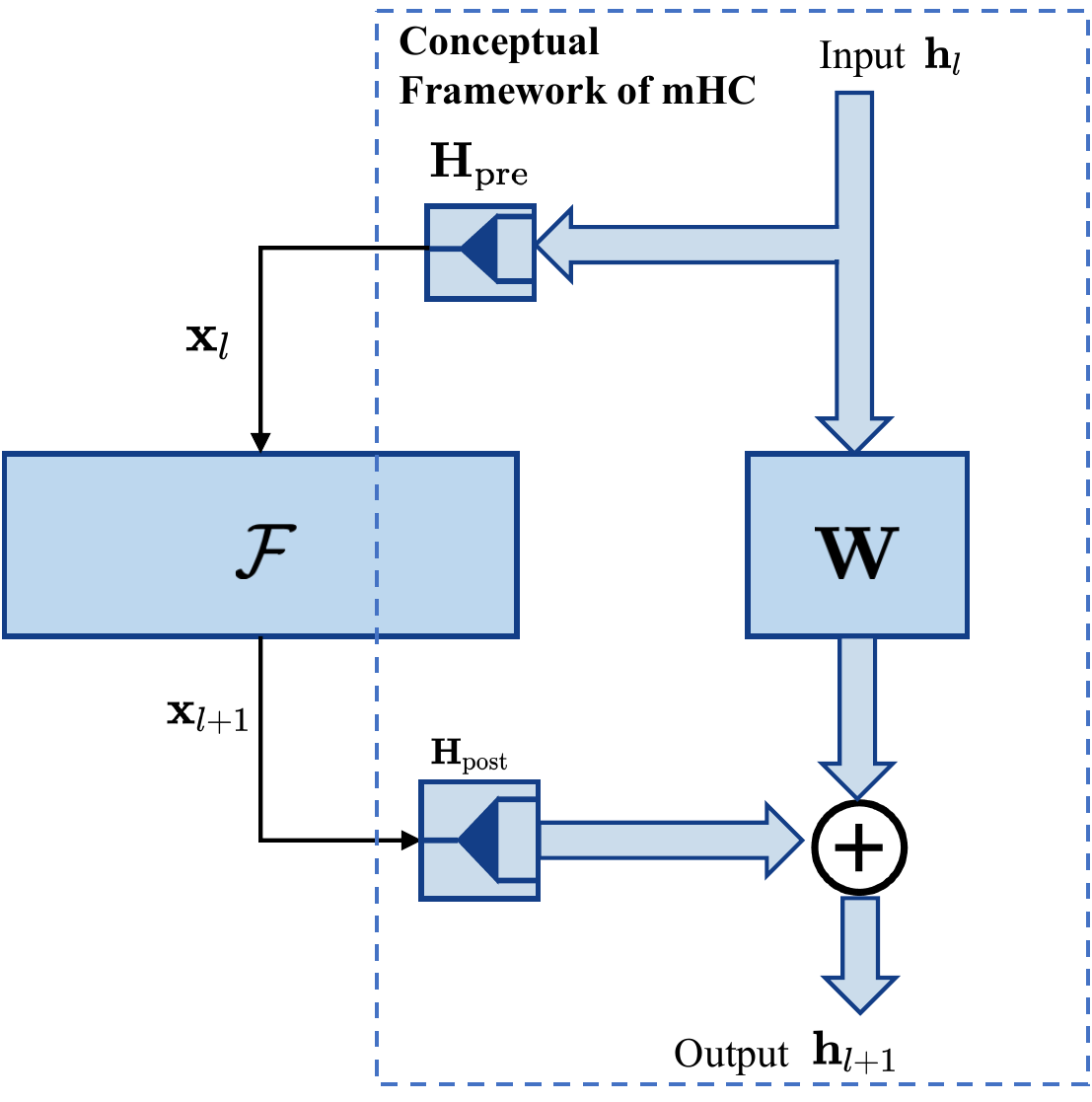} 
    \caption{High-level mHC}
    \label{fig:mhc}
  \end{subfigure}%
  \hspace{0.001\textwidth}
  \begin{subfigure}[b]{0.35\textwidth}
    \centering
    \includegraphics[height=5.0cm, width=!]{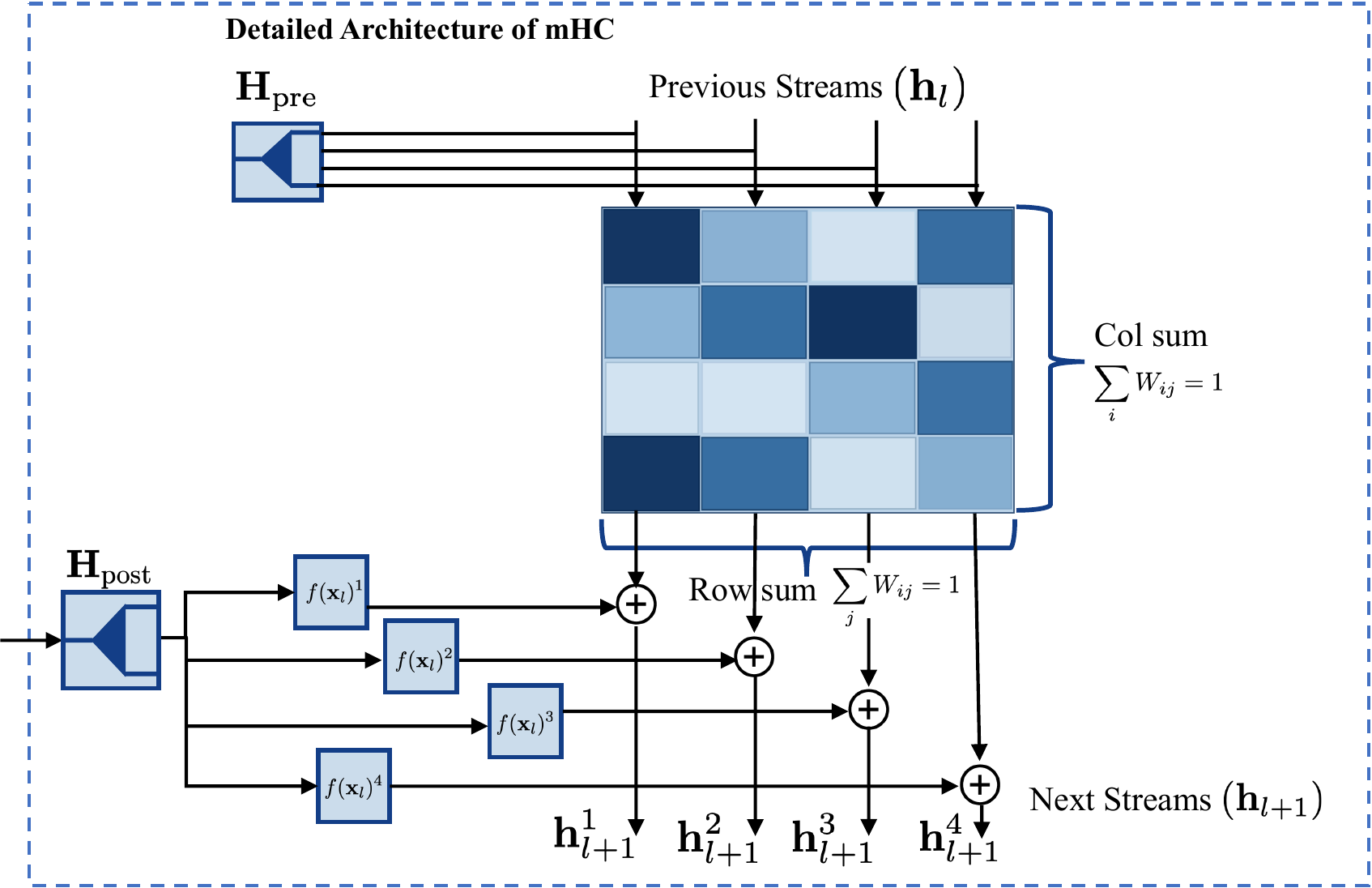}
    \caption{Detailed mHC}
    \label{fig:mhc_detail}
  \end{subfigure}

  \caption{Architectural comparison between the standard residual shortcut and mHC: (a) standard residual connection, (b) high-level mHC overview, and (c) detailed mHC design, where $\mathbf{H}_{\mathrm{pre}}$ aggregates streams before $\mathcal{F}(\cdot)$ and $\mathbf{H}_{\mathrm{post}}$ splits the output back into streams.}
  \label{fig:comparison}
\end{figure*}

\section{Proposed Method}
\subsection{Preliminaries: Manifold-Constrained Hyper-Connections}
Fig.~\ref{fig:residual} illustrates the standard residual connection. At layer $l$, the output is computed as
$\mathbf{x}_{l+1} = \mathbf{x}_l + \mathcal{F}(\mathbf{x}_l)$,
where $\mathbf{x}_l$ is passed through the shortcut unchanged and then added to the transformation block $\mathcal{F}(\mathbf{x}_l)$. This identity shortcut helps to train deep networks, but the shortcut itself is \emph{channel-wise independent}: the $p$-th channel of $\mathbf{x}_l$ is only added to the $p$-th channel of $\mathcal{F}(\mathbf{x}_l)$, so there is no explicit mechanism to redistribute information across channels within the skip path.

In Fig.~\ref{fig:mhc}, we use arrow widths to highlight the structural evolution: the thin arrows correspond to the single-path flow used in standard residual connections (Fig.~\ref{fig:residual}), while the wide arrows symbolize the multi-stream state of mHC, where information is processed in parallel subspaces. This visual distinction clarifies how mHC generalizes the rigid identity shortcut into a more expressive hyper-connection. Fig.~\ref{fig:mhc_detail} details the mHC architecture. The hidden state at layer $l$ is represented as $N$ parallel streams, $\mathbf{h}_{l} = \{\mathbf{h}_{l}^1, \mathbf{h}_{l}^2, \dots, \mathbf{h}_{l}^N\}$. These streams are first aggregated via an operator $\mathbf{H}_{\text{pre}}$ (concatenation) to form a unified feature map $\mathbf{x}_{l}$, which is then fed into the transformation block $\mathcal{F}(\cdot)$. The resulting output is subsequently partitioned by the operator $\mathbf{H}_{\text{post}}$ into $N$ disjoint streams $\{f(\mathbf{x}_l)^1, \dots, f(\mathbf{x}_l)^N\}$. This structure enables a stream-wise update where the transition from $\mathbf{h}_{l}$ to $\mathbf{h}_{l+1}$ is formalized as:

\begin{equation}
\mathbf{h}_{l+1}^i = \sum_{j=1}^{N} W_{ij}\,\mathbf{h}_{l}^j + f(\mathbf{x}_l)^i, \quad \forall i \in \{1, \dots, N\}.
\label{eq:mhc_stream_update}
\end{equation}
Inter-stream communication is governed by the learnable mixing matrix $\mathbf{W} \in \mathbb{R}^{N \times N}$, where $W_{ij}$ controls the information flow from the $j$-th historical stream to the $i$-th updated stream. This weighted summation provides the explicit mechanism for inter-channel communication missing in standard residuals. Unlike identity shortcuts where information is confined to isolated channels, $\mathbf{W}$ allows every stream $i$ to selectively aggregate and ``refresh" its content with all $N$ previous streams, transforming the shortcut into a high-bandwidth system for representation exchange.

To maintain numerical stability, $\mathbf{W}$ is projected onto a doubly stochastic manifold using Sinkhorn-Knopp iterations. We first construct a nonnegative matrix $\mathbf{A} = \exp(\mathbf{\Theta})$ from a learnable parameter matrix $\mathbf{\Theta}$. Then, we alternately apply row normalization $\mathbf{A} \leftarrow \text{diag}(\mathbf{A}\mathbf{1}_N)^{-1}\mathbf{A}$ followed by column normalization $\mathbf{A} \leftarrow \mathbf{A}\,\text{diag}(\mathbf{1}_N^\top\mathbf{A})^{-1}$. After $K$ iterations, we obtain the projection matrix $\mathbf{W}$ as the normalized matrix $\mathbf{A}$ that enforces unit row and column sums. This ``Energy Conservation" constraint preserves the signal scale across streams, preventing the signal from exploding or vanishing with depth. Compared with the standard residual shortcut, mHC enables active, learnable mixing while retaining the robust optimization properties of the original architecture.

In \cite{jin2026speaker_mhc}, we inserted mHC at the main residual shortcuts of the speaker encoders. Specifically,
it replaced the inter-block shortcuts in ResNet-34 and Res2Net and the main
shortcut of the SE-Res2Block in ECAPA-TDNN. These Fbank-based encoders were
trained end-to-end, allowing both the feature transformations and the mHC
connectivity to be jointly optimized.

\subsection{Manifold-Constrained Residual Adapter}
Most existing PEFT methods adapt the feature
transformations of a pretrained model, while leaving its residual connections
unchanged. However, residual connections determine how representations are
propagated across layers, and even a small modification can influence all
subsequent layers. Meanwhile, mHC provides an efficient mechanism for adapting
this information flow: its small $N\times N$ mixing matrix introduces very few
parameters while acting on the complete residual state. Motivated by these
observations, we propose the Manifold-Constrained Residual Adapter (MCRA),
which applies lightweight group-wise mixing to the residual paths of frozen
WavLM and uses a learnable gate to preserve the pretrained mapping at
initialization.

\begin{figure*}[!t]
    \centering
    \makebox[\textwidth][c]{%
        \includegraphics[width=1.0\textwidth]{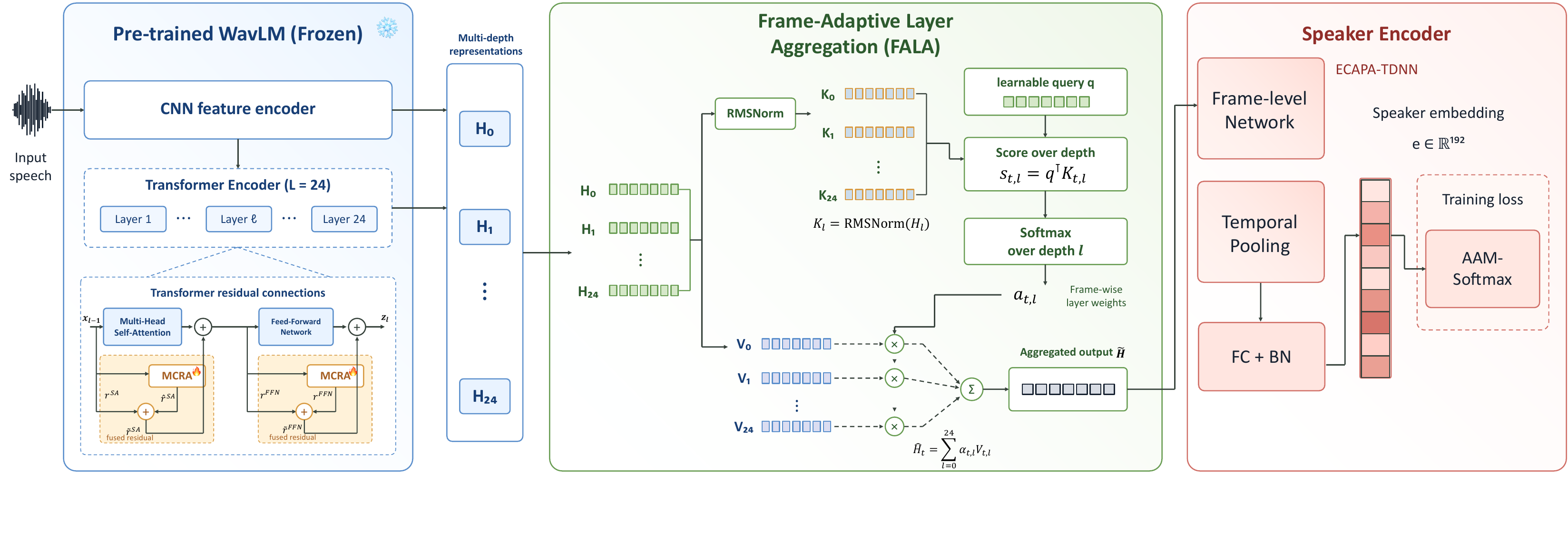}%
    }
    \caption{Overall architecture of the proposed system. The pretrained
    WavLM network is frozen. In each Transformer layer, MCRA produces the mixed
    residual $\widehat{\vect{r}}^{c}$, which is fused with the original
    residual $\vect{r}^{c}$ through a learnable gate to obtain
    $\widetilde{\vect{r}}^{c}$, where
    $c\in\{\mathrm{SA},\mathrm{FFN}\}$. FALA then computes frame-dependent
    layer weights before the speaker encoder. Frozen LayerNorm and dropout
    operations are omitted from the diagram for clarity.}
    \label{fig:overview}
\end{figure*}

As shown in the left part of Fig.~\ref{fig:overview}, MCRA is inserted into
both the self-attention (SA) and FFN residual paths of every WavLM Transformer
layer. Consider the SA residual path in layer $l$. Its input is
$\vect{x}_l$. Following the detailed mHC operation in
Fig.~\ref{fig:mhc_detail}, $\mathbf{H}_{\mathrm{post}}$ divides
$\vect{x}_l$ into $N$ feature groups, the matrix $\mat{W}$ mixes information
among these groups, and $\mathbf{H}_{\mathrm{pre}}$ combines them back into a
$D$-dimensional residual. A gate then blends this mixed residual with the
original $\vect{x}_l$. The resulting residual is added to the output of the
self-attention sublayer. MCRA performs the same operation on the FFN residual
path.

For a compact description of the two residual paths, we use
$c\in\{\mathrm{SA},\mathrm{FFN}\}$ to indicate the residual type and denote
its original residual state by $\vect{r}_l^c$. In the SA path,
$\vect{r}_l^{\mathrm{SA}}=\vect{x}_l$. Each layer and residual path has its own
$N\times N$ mixing matrix $\mat{W}_l^c$, obtained from learnable logits
$\mat{\Theta}_l^c$ using the same Sinkhorn normalization described in the
previous subsection. Applying $\mathbf{H}_{\mathrm{post}}$, $\mat{W}_l^c$,
and $\mathbf{H}_{\mathrm{pre}}$ in sequence implements the mixed shortcut term
in Eq.~\eqref{eq:mhc_stream_update} and produces the MCRA output
$\widehat{\vect{r}}_l^c$. This is the mixed residual shown in
Fig.~\ref{fig:overview}. The two $\mathbf{H}$ operators only partition and
recombine the feature groups and contain no trainable parameters.

To preserve the pretrained residual path at the beginning of adaptation,
MCRA learns one scalar gate logit $g_l^c$ for each residual junction. Its
sigmoid value $\lambda_l^c$ controls the interpolation between the original
and mixed residuals:
\begin{align}
    \lambda_l^c &=\sigma(g_l^c), \\
    \widetilde{\vect{r}}_l^c
        &=(1-\lambda_l^c)\vect{r}_l^c
          +\lambda_l^c\widehat{\vect{r}}_l^c.
    \label{eq:mcra_fusion}
\end{align}
In Fig.~\ref{fig:overview}, the MCRA box produces
$\widehat{\vect{r}}_l^c$, and the following fusion produces
$\widetilde{\vect{r}}_l^c$. We initialize $g_l^c=-6$, so
$\lambda_l^c\approx0.0025$. Because
$\widetilde{\vect{r}}_l^c-\vect{r}_l^c
=\lambda_l^c(\widehat{\vect{r}}_l^c-\vect{r}_l^c)$, only about $0.25\%$ of
the difference between the mixed and original residuals is introduced at
initialization. Together with the scale-preserving doubly stochastic mixer,
this keeps the initial residual path close to the pretrained one.

The WavLM-Large Transformer layer uses a pre-normalized structure in which the
SA block is followed by the FFN block. Let $\vect{u}_l$ denote the output of
the SA residual block. We use $\operatorname{LN}_l^{\mathrm{SA}}$ and
$\operatorname{LN}_l^{\mathrm{FFN}}$ to denote the frozen layer-normalization
modules preceding the SA and FFN sublayers, respectively. Omitting the frozen
dropout operations for clarity, the complete layer update is
\begin{align}
    \vect{u}_l
        &=\operatorname{SA}_l\!\left(
          \operatorname{LN}_l^{\mathrm{SA}}(\vect{x}_l)\right)
          +\widetilde{\vect{r}}_l^{\mathrm{SA}}, \\
    \vect{r}_l^{\mathrm{FFN}}&=\vect{u}_l, \\
    \vect{x}_{l+1}
        &=\operatorname{FFN}_l\!\left(
          \operatorname{LN}_l^{\mathrm{FFN}}(\vect{u}_l)\right)
          +\widetilde{\vect{r}}_l^{\mathrm{FFN}}.
    \label{eq:mcra_transformer_update}
\end{align}
Thus, the SA path applies MCRA to $\vect{x}_l$, while the FFN path applies the
same MCRA operation to $\vect{u}_l$.

\subsection{Frame-Adaptive Layer Aggregation}
Let $l\in\{0,\ldots,L\}$ index the representations extracted from WavLM. We
denote the $l$-th representation by
$\mat{X}_l=[\vect{x}_{l,1},\ldots,\vect{x}_{l,T}]^{\top}
\in\mathbb{R}^{T\times D}$, where $\mat{X}_0$ is the convolutional output and
$\mat{X}_l$ for $l\geq1$ is the output of the $l$-th Transformer layer. Here,
$\vect{x}_{l,t}\in\mathbb{R}^{D}$ denotes frame $t$ in $\mat{X}_l$. A
conventional learned layer-weighted sum is
\begin{align}
    \beta_l
        &=\frac{\exp(b_l)}
        {\sum_{j=0}^{L}\exp(b_j)},
        \label{eq:fixed_layer_weight}\\
    \mat{Z}_{\mathrm{fixed}}
        &=\sum_{l=0}^{L}\beta_l\mat{X}_l.
        \label{eq:fixed_layer_sum}
\end{align}
Here, $b_l$ is the learnable parameter for representation $l$, and $\beta_l$
is the corresponding normalized layer weight; $j$ runs over all $L+1$
representations in the
softmax denominator. The resulting matrix
$\mat{Z}_{\mathrm{fixed}}
=[\vect{z}^{\mathrm{fixed}}_1,\ldots,
\vect{z}^{\mathrm{fixed}}_T]^{\top}\in\mathbb{R}^{T\times D}$
is the aggregated feature sequence. Because $\beta_l$ depends only on the
layer index $l$, the same layer weights are used for every utterance and
frame.

FALA instead uses a single learnable query $\vect{q}\in\mathbb{R}^{D}$,
shared across all utterances and frames, to represent a task-specific
direction for speaker-relevant information. For a fixed frame $t$, FALA first
collects the representations from all layers,
$\{\vect{x}_{0,t},\ldots,\vect{x}_{L,t}\}$. Taking layer $l$ as an example,
$\vect{x}_{l,t}\in\mathbb{R}^{D}$ is its frame-level representation. Before
computing the layer attention, FALA applies
RMSNorm~\cite{zhang2019rmsnorm} to its $D$ feature channels:
\begin{align}
    \eta_{l,t}
        &=\sqrt{\frac{1}{D}\sum_{d=1}^{D}x_{l,t,d}^{2}+\epsilon}, \\
    \vect{k}_{l,t}
        &=\vect{\gamma}\odot\frac{\vect{x}_{l,t}}{\eta_{l,t}},
        \qquad \vect{\gamma}\in\mathbb{R}^{D}.
    \label{eq:fala_key}
\end{align}
Here, $x_{l,t,d}$ is channel $d$ of $\vect{x}_{l,t}$,
$\eta_{l,t}$ is its root-mean-square magnitude, $\epsilon$ is a small constant
for numerical stability, and $\odot$ denotes element-wise multiplication.
The vector $\vect{\gamma}$ is the learnable channel-wise scale of RMSNorm.

At frame $t$, the shared query scores the key from each layer. A softmax over
the layer index produces the layer weights:
\begin{align}
    a_{l,t}&=\frac{\vect{q}^{\top}\vect{k}_{l,t}}{\sqrt{D}}, \\
    \alpha_{l,t}&=\frac{\exp(a_{l,t})}
    {\sum_{j=0}^{L}\exp(a_{j,t})}, \\
    \vect{z}_t&=\sum_{l=0}^{L}\alpha_{l,t}\vect{x}_{l,t}, \\
    \mat{Z}&=[\vect{z}_1,\ldots,\vect{z}_T]^{\top}
    \in\mathbb{R}^{T\times D}.
    \label{eq:fala}
\end{align}
Here, $\alpha_{l,t}$ is the weight of layer $l$ at frame $t$,
$\vect{z}_t\in\mathbb{R}^{D}$ is the layer-aggregated representation of that
frame, and $\mat{Z}$ is the complete FALA output. Since the keys depend on the
input, the layer weights can vary across utterances and frames.

\subsection{Parameter Efficiency}
MCRA and FALA adapt the information flow of WavLM while keeping all original
encoder parameters frozen. Their parameter costs are determined solely by
the proposed residual mixers, gates, query, and normalization scale.

In particular, MCRA assigns an $N\times N$ mixing matrix and one scalar gate to both the SA
and FFN residual junctions in every Transformer layer. Its parameter count is
\begin{equation}
    P_{\mathrm{MCRA}}=2L(N^2+1).
    \label{eq:mcra_parameters}
\end{equation}
On the other hand, FALA contains no query, key, or value (Q/K/V) projection matrices used in
standard Transformer attention~\cite{vaswani2017attention}, multi-head
transformation, attention MLP, or output projection. Its only trainable
parameters are the
$D$-dimensional query $\vect{q}$ and the $D$-dimensional RMSNorm scale
$\vect{\gamma}$, giving
\begin{equation}
    P_{\mathrm{FALA}}=2D.
    \label{eq:fala_parameters}
\end{equation}
Therefore, the total number of WavLM-side adaptation parameters is
\begin{align}
    P_{\mathrm{adapt}}
        &=P_{\mathrm{MCRA}}+P_{\mathrm{FALA}} \\
        &=2L(N^2+1)+2D.
    \label{eq:total_adaptation_parameters}
\end{align}
For WavLM-Large with $L=24$, $N=4$, and $D=1024$,
$P_{\mathrm{MCRA}}=816$ and $P_{\mathrm{FALA}}=2{,}048$, giving
$P_{\mathrm{adapt}}=2{,}864$, demonstrating that the proposed
parameter-efficient adaptation requires very few learnable parameters.

\section{Experimental Setup}
\subsection{Datasets}
\textbf{VoxCeleb:} We use the VoxCeleb2 development set for training, which
contains 1,092,009 utterances from 5,994 speakers~\cite{chung2018voxceleb2}.
The systems are evaluated on three VoxCeleb1 trial lists: VoxCeleb1-O with
37,720 trials, VoxCeleb1-E with 581,480 trials, and VoxCeleb1-H with 552,536
trials~\cite{nagrani2017voxceleb}. We additionally use the 60,000 trials in
VoxSRC21-val~\cite{brown2022voxsrc} for robustness evaluation.

\textbf{CN-Celeb:} CN-Celeb is an in-the-wild Chinese speaker recognition
corpus containing 659,594 utterances from 3,000 speakers across 11
genres~\cite{li2022cnceleb}. The systems are trained on the training portions
of CN-Celeb1 and CN-Celeb2 and evaluated on CN-Celeb1.Eval, which contains
200 speakers.

\textbf{Data augmentation resources:} MUSAN~\cite{snyder2015musan} provides
the noise, speech, and music recordings used for additive augmentation. Room
impulse responses from the RIR dataset~\cite{ko2017reverberant} are used to
generate reverberated speech.

\subsection{Implementation Details}
Each training sample contains 400 frames, corresponding to 64,240 waveform
samples, or approximately 4.015~s at 16~kHz. The batch size is 128. SGD is
used with an initial learning rate of 0.002, momentum of 0.9, and weight
decay of $10^{-4}$. A StepLR scheduler decays the learning rate by a factor
of 0.97 after each epoch. The 5,994-way classification head is trained using
AAM-Softmax~\cite{deng2019arcface} with margin $m=0.2$ and scale $s=30$.

Each training utterance is augmented with probability 0.6. One augmentation
type is randomly selected from reverberation, noise, speech, music, and
speech-plus-music. For MUSAN augmentation, the SNR ranges are 0--15~dB for
noise, 13--20~dB for speech, and 5--15~dB for music. We further apply
SpecAugment to the input features of ECAPA-TDNN, with maximum
frequency- and time-mask widths of 8 and 10, respectively.

\subsection{Model Configurations}
We use WavLM-Base+ and WavLM-Large as the backbones~\cite{chen2022wavlm}.
WavLM-Base+ contains 12 Transformer layers with a hidden dimension of 768,
whereas WavLM-Large contains 24 Transformer layers with a hidden dimension
of 1,024. Both backbones are followed by an ECAPA-TDNN with 1,024
channels~\cite{desplanques2020ecapa}, which produces a 192-dimensional
speaker embedding.

All WavLM parameters are frozen, while the ECAPA-TDNN backend, AAM-Softmax
classifier, MCRA, and FALA are trainable.

Unless otherwise stated, MCRA uses $N=4$ feature groups, eight Sinkhorn
normalization iterations, and a learnable gate initialized with $g_l^c=-6$.
MCRA is applied to both the self-attention and FFN residual junctions.

\subsection{Scoring and Evaluation Metrics}
Cosine similarity between $\ell_2$-normalized speaker embeddings is used for
scoring. Performance is evaluated using equal error rate (EER) and minimum
detection cost function (minDCF), with $P_{\mathrm{target}}=0.01$,
$C_{\mathrm{miss}}=1$, and $C_{\mathrm{fa}}=1$. For AS-Norm, the cohort
contains 6,000 speakers with five utterances per speaker, and the top 300
cohort scores are retained.

\section{Results and Analysis}
\subsection{Generality of Manifold-Constrained Mixing}
Tables~\ref{tab:previous_mhc} and~\ref{tab:cnceleb_results} examine whether
the benefit of mHC depends on a particular speaker encoder or evaluation
set. Table~\ref{tab:previous_mhc} summarizes the results of our preliminary
Interspeech study~\cite{jin2026speaker_mhc}. Replacing standard residual
connections with mHC reduces EER for all four evaluated backbones: Res2Net,
ResNet-34, ECAPA-TDNN-S, and ECAPA-TDNN-L. These models use different
network structures and range from 4.03M to 20.76M parameters. Their
consistent improvements indicate that manifold-constrained mixing is not
specific to one architecture or model size.

The EER improvements also hold on VoxCeleb1-O, E, and H, as well as
VoxSRC21-val. In particular, mHC remains effective on VoxCeleb1-H and
VoxSRC21-val, where all systems have higher error rates than on
VoxCeleb1-O. This shows that its benefit is not limited to the original
VoxCeleb1 trial list and remains visible under more challenging evaluation
conditions. ECAPA-TDNN-L, for example, reduces EER from 0.87\%, 1.12\%, and
2.12\% to 0.77\%, 0.94\%, and 1.88\% on VoxCeleb1-O, E, and H, respectively,
and from 3.89\% to 3.68\% on VoxSRC21-val.

To further validate the effectiveness of mHC, we also evaluate it on the
Chinese CN-Celeb dataset, as shown in Table~\ref{tab:cnceleb_results}. For
ECAPA-TDNN-L, mHC reduces EER from 8.65\% to 8.01\%
and minDCF from 0.498 to 0.432, corresponding to relative reductions of
7.4\% and 13.3\%, respectively. It also achieves the best EER and minDCF
among the listed Fbank systems. This result suggests that the improvement
extends beyond the English VoxCeleb domain to a different language and
acoustic domain. Overall, these results demonstrate the generality of mHC
across architectures, model sizes, evaluation conditions, and domains.

\begin{table*}[t]
\caption{Comparison with representative Fbank-based speaker verification systems, including recent state-of-the-art methods. EER (\%) and minDCF are reported on the standard VoxCeleb1 test sets and VoxSRC21-val. The best result within each architecture group is shown in bold.}
\label{tab:previous_mhc}
\centering
\setlength{\tabcolsep}{5pt}
\renewcommand{\arraystretch}{1.06}
\resizebox{0.9\textwidth}{!}{%
\begin{tabular}{l|c|c|c|c|c|c|c|c|c}
\hline
\multirow{2}{*}{Architecture}
& \multirow{2}{*}{Params (M)}
& \multicolumn{2}{c|}{VoxCeleb1-O}
& \multicolumn{2}{c|}{VoxCeleb1-E}
& \multicolumn{2}{c|}{VoxCeleb1-H}
& \multicolumn{2}{c}{VoxSRC21-val} \\
\cline{3-10}
&
& EER(\%) & minDCF
& EER(\%) & minDCF
& EER(\%) & minDCF
& EER(\%) & minDCF \\
\hline

Res2Net-48w2s~\cite{zhou2021resnext_res2net}
& 5.50
& 1.75 & 0.183
& 1.70 & 0.182
& 2.98 & 0.283
& -- & -- \\

Res2Net-MWA~\cite{gu2023memory,chen2023eres2net}
& 5.50
& 1.71 & 0.165
& -- & --
& -- & --
& -- & -- \\

Res2Net
& 4.03
& 1.56 & \textbf{0.150}
& 1.41 & 0.149
& 2.48 & 0.230
& 4.27 & 0.334 \\

mHC-Res2Net~\cite{jin2026speaker_mhc}
& 4.03
& \textbf{1.41} & 0.154
& \textbf{1.40} & \textbf{0.145}
& \textbf{2.43} & \textbf{0.228}
& \textbf{4.13} & \textbf{0.332} \\

\hline

ResNet34-FPM~\cite{jung2020fpm,chen2023eres2net}
& 5.85
& 1.98 & 0.205
& -- & --
& -- & --
& -- & -- \\

ResNet34-SKDFE~\cite{liu2022skdfe}
& 5.98
& 1.44 & 0.168
& 1.59 & 0.179
& 2.76 & 0.278
& -- & -- \\

Res2Former-Base~\cite{chen2025res2former}
& 1.73
& 1.23 & 0.120
& 1.36 & 0.140
& 2.38 & 0.230
& -- & -- \\

ResNet-34
& 6.34
& 1.05 & 0.107
& 1.11 & 0.116
& 1.99 & 0.192
& 3.83 & 0.316 \\

mHC-ResNet-34~\cite{jin2026speaker_mhc}
& 6.34
& \textbf{1.03} & \textbf{0.095}
& \textbf{1.09} & \textbf{0.110}
& \textbf{1.93} & \textbf{0.188}
& \textbf{3.35} & \textbf{0.269} \\

\hline

TDNN~\cite{wang2023campp}
& 4.62
& 2.31 & 0.322
& 2.37 & 0.273
& 4.25 & 0.393
& -- & -- \\

D-TDNN~\cite{wang2023campp}
& 2.85
& 1.55 & 0.166
& 1.63 & 0.175
& 2.86 & 0.257
& -- & -- \\

ECAPA-TDNN-S
& 6.19
& 1.02 & \textbf{0.106}
& 1.41 & 0.162
& 2.26 & 0.257
& 4.26 & 0.384 \\

mHC-ECAPA-TDNN-S~\cite{jin2026speaker_mhc}
& 6.19
& \textbf{0.98} & 0.107
& \textbf{1.07} & \textbf{0.119}
& \textbf{2.06} & \textbf{0.204}
& \textbf{4.11} & \textbf{0.324} \\

\hline

E-TDNN-L~\cite{desplanques2020ecapa}
& 20.40
& 1.26 & 0.140
& 1.37 & 0.149
& 2.35 & 0.215
& -- & -- \\

D-TDNN-L~\cite{wang2023campp}
& 6.40
& 1.19 & 0.118
& 1.21 & 0.129
& 2.22 & 0.205
& -- & -- \\

MGFF-TDNN~\cite{li2025mgfftdnn}
& 4.78
& 0.89 & 0.086
& 1.05 & 0.119
& 1.91 & 0.195
& -- & -- \\

ECAPA-TDNN-L
& 20.76
& 0.87 & 0.107
& 1.12 & 0.132
& 2.12 & 0.210
& 3.89 & 0.323 \\

mHC-ECAPA-TDNN-L~\cite{jin2026speaker_mhc}
& 20.76
& \textbf{0.77} & \textbf{0.084}
& \textbf{0.94} & \textbf{0.110}
& \textbf{1.88} & \textbf{0.190}
& \textbf{3.68} & \textbf{0.316} \\

\hline
\end{tabular}%
}
\end{table*}

\begin{table}[t]
\caption{Results on the CN-Celeb evaluation set.}
\label{tab:cnceleb_results}
\centering
\small
\setlength{\tabcolsep}{6pt}
\renewcommand{\arraystretch}{1.06}
\resizebox{\columnwidth}{!}{%
\begin{tabular}{lccc}
\toprule
Architecture & Params (M) & EER (\%) & minDCF \\
\midrule

ECAPA-TDNN-L (ours)~\cite{desplanques2020ecapa}
& 20.76 & 8.65 & 0.498 \\

ECAPA-TDNN+AAMSupCon~\cite{li2022aamsupcon}
& -- & 8.49 & 0.500 \\

D-TDNN~\cite{wang2023campp}
& 2.85 & 8.41 & 0.468 \\

SC-EcapaTdnn~\cite{zhang2024scecapa}
& 8.28 & 8.26 & 0.480 \\

\midrule

mHC-ECAPA-TDNN-L (ours), $G=8$
& 20.76 & \textbf{8.01} & \textbf{0.432} \\

\bottomrule
\end{tabular}
}
\end{table}

\subsection{Comparison with Existing Adaptation Methods}
Table~\ref{tab:adaptation_comparison} compares the proposed methods with
representative full fine-tuning and PEFT approaches on VoxCeleb1. The
proposed methods are evaluated with both WavLM-Base+ and WavLM-Large.
The MCRA, FALA, and MCRA--FALA results within each WavLM group use the same
training and evaluation setup. 

\begin{table*}[t]
\caption{Comparison with representative fine-tuning and parameter-efficient
adaptation methods on VoxCeleb1. The AS-Norm setting follows each source
paper. Full fine-tuning results are shown in italics, and the best PEFT
results for each PTM and metric are highlighted in bold.}
\label{tab:adaptation_comparison}
\centering
\setlength{\tabcolsep}{3.2pt}
\renewcommand{\arraystretch}{1.10}
\resizebox{\textwidth}{!}{%
\begin{tabular}{l|l|l|c|c|c|c|c|c|c|c|c}
\hline
\multirow{2}{*}{PTM}
& \multirow{2}{*}{Encoder}
& \multirow{2}{*}{Fine-tuning Method}
& \multirow{2}{*}{AS-Norm}
& \multirow{2}{*}{\shortstack{Per-Module\\\#Params}}
& \multirow{2}{*}{\shortstack{Trainable\\\#Params in PTM}}
& \multicolumn{2}{c|}{VoxCeleb1-O}
& \multicolumn{2}{c|}{VoxCeleb1-E}
& \multicolumn{2}{c}{VoxCeleb1-H} \\
\cline{7-12}
& & & & & &
EER(\%) & minDCF &
EER(\%) & minDCF &
EER(\%) & minDCF \\
\hline
\multirow{5}{*}{WavLM-Base+}
& \multirow{5}{*}{TDNN}
& LoRA \cite{hu2022lora,sang2025unipet}
& \xmark
& 2.40M
& 2.40M
& 2.35 & 0.179
& 2.65 & 0.178
& 5.04 & 0.288 \\
\cline{3-12}
& & Houlsby Adapter \cite{houlsby2019parameter,sang2025unipet}
& \xmark
& 9.50M
& 9.50M
& 2.30 & 0.202
& 2.77 & 0.204
& 5.60 & 0.339 \\
\cline{3-12}
& & Inner+Inter Adapter \cite{sang2025unipet}
& \xmark
& 4.80M
& 4.80M
& 1.82 & 0.156
& 2.18 & 0.150
& 4.05 & 0.251 \\
\cline{3-12}
& & Deep Speaker Prompting \cite{sang2025unipet}
& \xmark
& 0.30M
& 0.30M
& 2.16 & 0.172
& 2.41 & 0.169
& 4.62 & 0.292 \\
\cline{3-12}
& & UniPET-SPK \cite{sang2025unipet}
& \xmark
& 5.10M
& 5.10M
& 1.59 & 0.130
& 1.74 & 0.138
& 3.41 & 0.206 \\
\hline
\multirow{7}{*}{WavLM-Base+}
& \multirow{7}{*}{ECAPA-TDNN}
& None (Frozen PTM) \cite{li2024dualpeft}
& \xmark
& 0.00M
& 0.00M
& 1.38 & 0.160
& 1.56 & 0.182
& 3.29 & 0.314 \\
\cline{3-12}
& & \multirow{2}{*}{MCRA}
& \xmark
& \multirow{2}{*}{0.017K}
& \multirow{2}{*}{0.421K}
& 1.29 & 0.140
& 1.32 & 0.133
& 2.48 & 0.229 \\
& & & \cmark & & &
1.17 & 0.120
& 1.18 & \textbf{0.118}
& 2.27 & 0.205 \\
\cline{3-12}
& & \multirow{2}{*}{FALA}
& \xmark
& \multirow{2}{*}{1.536K}
& \multirow{2}{*}{1.536K}
& 1.25 & 0.130
& 1.35 & 0.135
& 2.48 & 0.226 \\
& & & \cmark & & &
1.11 & 0.115
& 1.23 & \textbf{0.118}
& 2.20 & 0.200 \\
\cline{3-12}
& & \multirow{2}{*}{MCRA + FALA}
& \xmark
& \multirow{2}{*}{1.553K}
& \multirow{2}{*}{1.944K}
& 1.16 & 0.110
& 1.27 & 0.125
& 2.27 & 0.209 \\
& & & \cmark & & &
\textbf{1.04} & \textbf{0.100}
& \textbf{1.12} & 0.119
& \textbf{2.04} & \textbf{0.185} \\
\hline
\multirow{10}{*}{WavLM-Large}
& \multirow{10}{*}{ECAPA-TDNN}
& Full fine-tuning \cite{li2026unifiedpeft}
& \cmark
& --
& 316.00M
& \textit{0.38} & \textit{0.057}
& \textit{0.48} & \textit{0.086}
& \textit{0.98} & \textit{0.169} \\
\cline{3-12}
& & None (Frozen PTM) \cite{li2026unifiedpeft}
& \cmark
& 0.00M
& 0.00M
& 0.85 & 0.096
& 1.12 & 0.091
& 2.06 & 0.197 \\
\cline{3-12}
& & Adapter \cite{peng2023adapter,li2026unifiedpeft}
& \cmark
& 0.53M
& 12.61M
& 0.68 & 0.070
& 0.77 & 0.091
& 1.79 & 0.180 \\
\cline{3-12}
& & LoRA \cite{hu2022lora,li2026unifiedpeft}
& \cmark
& 0.08M
& 1.97M
& \textbf{0.65} & \textbf{0.065}
& \textbf{0.74} & \textbf{0.087}
& 1.72 & 0.174 \\
\cline{3-12}
& & \multirow{2}{*}{MCRA}
& \xmark
& \multirow{2}{*}{0.017K}
& \multirow{2}{*}{0.841K}
& 0.82 & 0.122
& 0.91 & 0.100
& 1.87 & 0.192 \\
& & & \cmark & & &
0.81 & 0.122
& 0.85 & 0.092
& 1.75 & 0.179 \\
\cline{3-12}
& & \multirow{2}{*}{FALA}
& \xmark
& \multirow{2}{*}{2.048K}
& \multirow{2}{*}{2.048K}
& 0.82 & 0.125
& 0.92 & 0.101
& 1.93 & 0.192 \\
& & & \cmark & & &
0.78 & 0.107
& 0.86 & 0.094
& 1.76 & 0.180 \\
\cline{3-12}
& & \multirow{2}{*}{MCRA + FALA}
& \xmark
& \multirow{2}{*}{2.065K}
& \multirow{2}{*}{2.864K}
& 0.72 & 0.090
& 0.91 & 0.102
& 1.81 & 0.175 \\
& & & \cmark & & &
0.66 & 0.085
& 0.84 & 0.092
& \textbf{1.69} & \textbf{0.168} \\
\hline
\end{tabular}%
}
\end{table*}

The proposed method improves the frozen baseline at both WavLM scales. For
WavLM-Base+, MCRA--FALA uses only 1.944K trainable PTM parameters. Without
AS-Norm, it reduces the EER from 1.38\%, 1.56\%, and 3.29\% to 1.16\%,
1.27\%, and 2.27\% on VoxCeleb1-O, E, and H, respectively. These results
correspond to relative EER reductions of 15.9\%, 18.6\%, and 31.0\%. The
consistent gains show that the proposed adaptation is effective for the
smaller WavLM-Base+ model.

The same trend is observed with WavLM-Large. With AS-Norm, MCRA--FALA
reduces the EER of the frozen baseline from 0.85\%, 1.12\%, and 2.06\% to
0.66\%, 0.84\%, and 1.69\%. The relative reductions are 22.4\%, 25.0\%, and
18.0\%, respectively. The minDCF also decreases from 0.096 to 0.085 on
VoxCeleb1-O and from 0.197 to 0.168 on VoxCeleb1-H. On VoxCeleb1-E, the
proposed system obtains a minDCF of 0.092, compared with 0.091 for the
frozen baseline. Overall, the consistent EER reductions across all three
trial lists demonstrate the effectiveness of the proposed system.

MCRA--FALA offers a favorable comparison with larger PEFT methods. For
WavLM-Large, it trains only 2.864K PTM parameters. This is approximately
688 times fewer than LoRA with 1.97M parameters and 4,403 times fewer than
the adapter with 12.61M parameters. It also uses over 110,000 times fewer
parameters than full fine-tuning. Despite this large reduction, its EER on
VoxCeleb1-O is 0.66\%, close to the 0.65\% obtained by LoRA. On
VoxCeleb1-H, MCRA--FALA outperforms LoRA in both EER and minDCF, reducing
them from 1.72\% and 0.174 to 1.69\% and 0.168. Its VoxCeleb1-E EER of
0.84\% remains higher than the 0.74\% achieved by LoRA. These results show
that MCRA--FALA retains competitive accuracy with substantially fewer
trainable parameters.

The WavLM-Base+ comparison shows a similar efficiency advantage, although
the published systems use a TDNN rather than our ECAPA-TDNN backend. The
1.944K parameters of MCRA--FALA are approximately 1,235 times fewer than
the 2.40M parameters of LoRA and 2,623 times fewer than the 5.10M parameters
of UniPET-SPK. Without AS-Norm, MCRA--FALA obtains EERs of 1.16\%, 1.27\%,
and 2.27\%, compared with 1.59\%, 1.74\%, and 3.41\% for UniPET-SPK. This
comparison demonstrates that MCRA--FALA maintains strong verification
performance while reducing the number of trainable PTM parameters by
several orders of magnitude.

The individual-module results provide an ablation of the proposed design.
On WavLM-Base+, both MCRA and FALA improve the frozen baseline. With
AS-Norm, MCRA obtains EERs of 1.17\%, 1.18\%, and 2.27\%, while FALA obtains
1.11\%, 1.23\%, and 2.20\%. Their combination further reduces the EERs to
1.04\%, 1.12\%, and 2.04\%. Relative to the better individual result on each
test set, the reductions are 6.3\%, 5.1\%, and 7.3\%. This confirms that
residual adaptation and frame-adaptive layer aggregation capture
complementary information.

The ablation results on WavLM-Large lead to the same conclusion. With
AS-Norm, MCRA achieves EERs of 0.81\%, 0.85\%, and 1.75\%, and FALA achieves
0.78\%, 0.86\%, and 1.76\%. MCRA--FALA further reduces these values to
0.66\%, 0.84\%, and 1.69\%. Compared with the better individual module, the
relative EER reductions are 15.4\%, 1.2\%, and 3.4\%. MCRA is slightly
better on VoxCeleb1-E and H, whereas FALA is better on VoxCeleb1-O. Neither
module therefore dominates across all test sets. Their combination gives
the strongest overall performance, demonstrating the benefit of adapting
both residual information propagation and layer aggregation.

\subsection{Cross-Domain Evaluation on CN-Celeb}
Table~\ref{tab:cnceleb_adaptation} evaluates the proposed methods on
CN-Celeb1.Eval under the CN-Celeb1\&2 training setting. This experiment tests
whether the trends observed on VoxCeleb remain valid in a different language
and acoustic domain.

\begin{table*}[t]
\caption{Comparison of the proposed parameter-efficient adaptation methods
on CN-Celeb1.Eval under the CN-Celeb1\&2 training setting. All MCRA
configurations use $N=8$. Best results are shown in bold.}
\label{tab:cnceleb_adaptation}
\centering
\setlength{\tabcolsep}{7pt}
\renewcommand{\arraystretch}{1.15}

\resizebox{0.82\textwidth}{!}{%
\begin{tabular}{l|l|l|c|c|c|c}
\hline
\multirow{2}{*}{PTM}
& \multirow{2}{*}{Encoder}
& \multirow{2}{*}{Fine-tuning Method}
& \multirow{2}{*}{\shortstack{Per-Module\\\#Params}}
& \multirow{2}{*}{\shortstack{Trainable\\\#Params in PTM}}
& \multicolumn{2}{c}{CN-Celeb1.Eval} \\
\cline{6-7}
& & & & & EER(\%) & minDCF \\
\hline

\multirow{5}{*}{WavLM-Large}
& \multirow{5}{*}{ECAPA-TDNN}
& None (Frozen PTM) \cite{li2026unifiedpeft}
& 0.00M
& 0.00M
& 8.72 & 0.448 \\
\cline{3-7}

& & Full fine-tuning \cite{li2026unifiedpeft}
& --
& 316.00M
& \textit{7.25} & \textit{0.428} \\
\cline{3-7}

& & MCRA
& 0.065K
& 3.145K
& 9.05 & 0.454 \\
\cline{3-7}

& & FALA
& 2.048K
& 2.048K
& 8.90 & 0.459 \\
\cline{3-7}

& & MCRA + FALA
& 2.113K
& 5.168K
& \textbf{8.47} & \textbf{0.447} \\
\hline

\end{tabular}%
}

\end{table*}

The proposed method also improves speaker verification performance on
CN-Celeb. With only a few thousand additional trainable parameters,
MCRA--FALA still improves the frozen WavLM baseline. Consistent improvements are observed with WavLM-Large, showing that the method remains effective. Together with the gains on VoxCeleb,
these results further demonstrate the robustness of the proposed method across model sizes, languages, and acoustic conditions.

\subsection{Effect of the Number of Feature Groups}
\begin{figure}[!t]
    \centering
    \begin{minipage}[t]{0.96\linewidth}
        \centering
        \includegraphics[width=\linewidth]{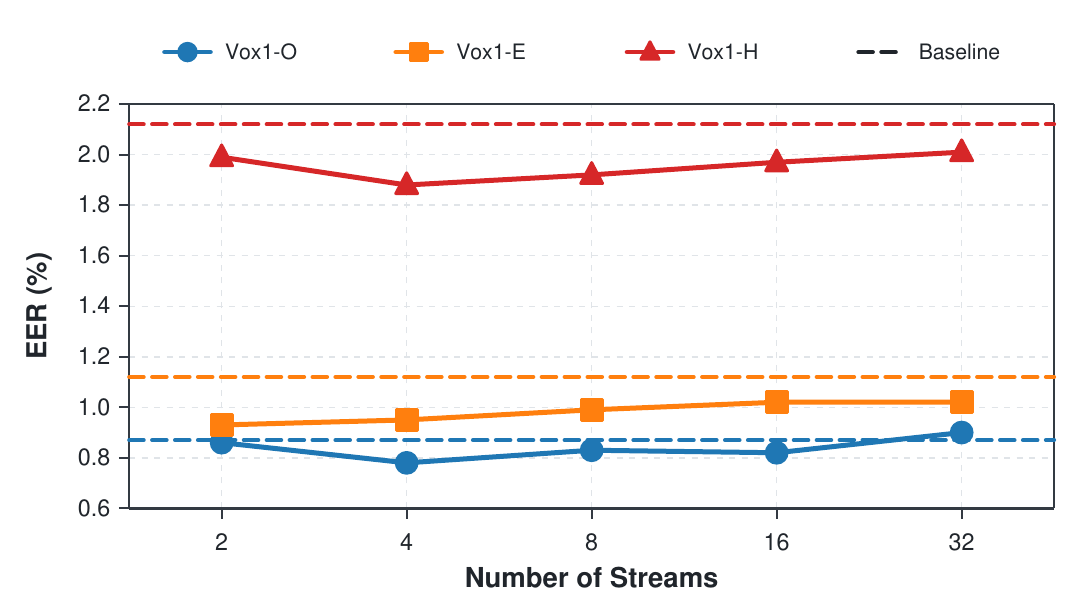}
        \small (a) VoxCeleb
    \end{minipage}
    \vspace{3pt}

    \begin{minipage}[t]{0.96\linewidth}
        \centering
        \includegraphics[width=\linewidth]{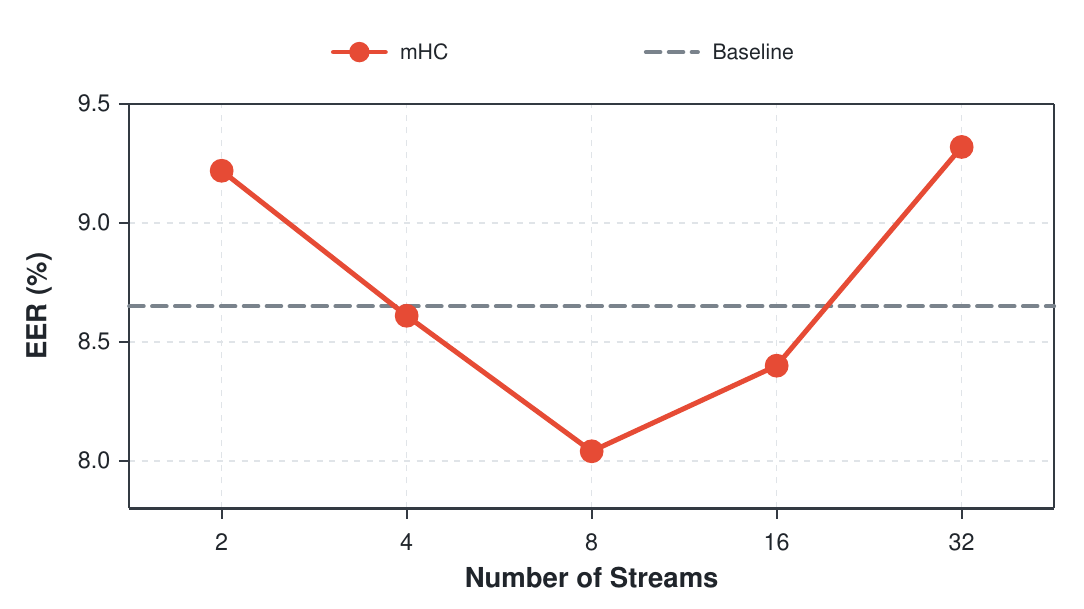}
        \small (b) CN-Celeb
    \end{minipage}
    \caption{Effect of the number of residual streams in the original mHC
    systems on (a) the VoxCeleb test sets and (b) CN-Celeb.}
    \label{fig:mhc_num_streams}
\end{figure}

\begin{figure}[!t]
    \centering
    \begin{minipage}[t]{0.96\linewidth}
        \centering
        \includegraphics[width=\linewidth]{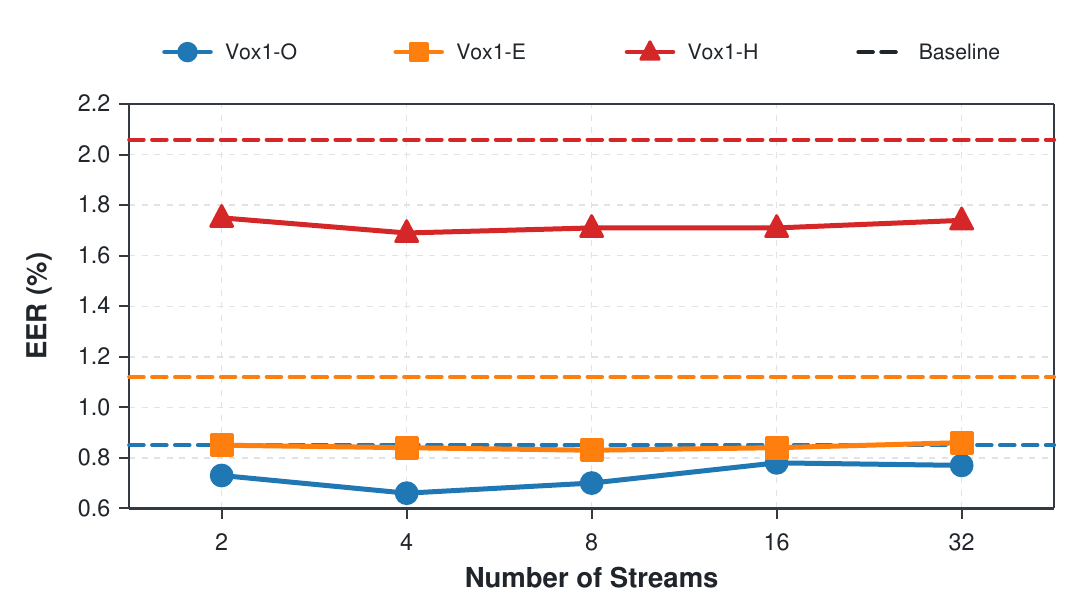}
        \small (a) VoxCeleb1
    \end{minipage}
    \vspace{3pt}

    \begin{minipage}[t]{0.96\linewidth}
        \centering
        \includegraphics[width=\linewidth]{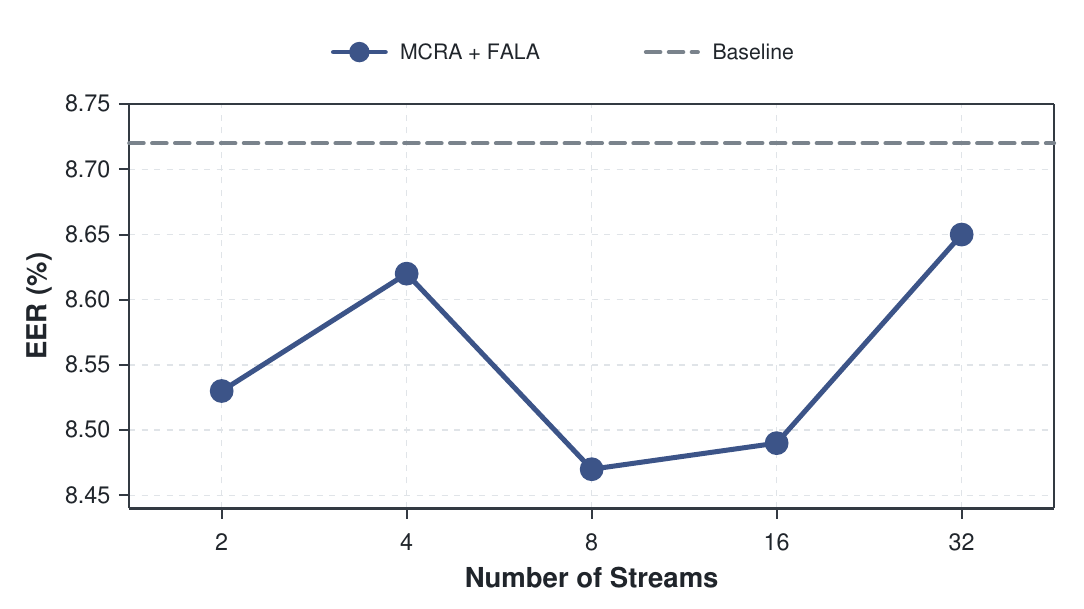}
        \small (b) CN-Celeb
    \end{minipage}
    \caption{Effect of the number of feature groups in the proposed
    MCRA--FALA system on (a) the VoxCeleb1 test sets and (b) CN-Celeb.}
    \label{fig:mcra_fala_num_groups}
\end{figure}

We further investigate the effect of varying the number of feature groups
$N$ on system performance, as shown in
Figs.~\ref{fig:mhc_num_streams} and~\ref{fig:mcra_fala_num_groups}.

For the original mHC experiments, we use ECAPA-TDNN-L as the speaker
encoder. $N=4$ achieves the best results on VoxCeleb1-O and VoxCeleb1-H,
while $N=2$ performs best on VoxCeleb1-E; on CN-Celeb, $N=8$ achieves the
lowest EER. For MCRA--FALA, we use WavLM-Large with an ECAPA-TDNN backend.
$N=4$ achieves the best results on VoxCeleb1-O and VoxCeleb1-H, whereas
$N=8$ performs best on VoxCeleb1-E and CN-Celeb. Accordingly, we use $N=4$
for the VoxCeleb experiments and $N=8$ for the CN-Celeb experiments.

\subsection{Comparison of Residual Mixing Constraints}
To verify the effectiveness of the Sinkhorn-based doubly stochastic
constraint in MCRA, we compare it with identity, unconstrained, and
row-softmax residual mixing using WavLM-Large with an ECAPA-TDNN backend.
MCRA uses $N=4$ feature groups and is applied to both the
self-attention and FFN residual connections; all other settings are kept
fixed. Table~\ref{tab:residual_mixing_constraints} reports the results on
VoxCeleb1-O.

\begin{table}[t]
\caption{Comparison of residual mixing constraints on VoxCeleb1-O.}
\label{tab:residual_mixing_constraints}
\centering
\setlength{\tabcolsep}{4.5pt}
\renewcommand{\arraystretch}{1.12}
\begin{tabular}{l|c|c|c|c}
\hline
Residual mixing & Row sum & Column sum & EER (\%) & minDCF \\
\hline
Identity & 1 & 1 & 0.82 & 0.114 \\
Unconstrained & -- & -- & 0.92 & 0.117 \\
Row softmax & 1 & -- & 0.88 & 0.114 \\
Sinkhorn & 1 & 1 & \textbf{0.66}& \textbf{0.085} \\
\hline
\end{tabular}
\end{table}

The identity setting uses a fixed, non-learnable identity matrix and
therefore performs no information exchange among feature groups. The unconstrained setting learns
the mixing matrix directly without controlling either its row or column
sums. Row softmax normalizes each row to sum to one, but leaves the column
sums unconstrained. Sinkhorn normalization alternately normalizes the rows
and columns, producing a doubly stochastic mixing matrix. Unconstrained
mixing and row softmax yield higher EERs than the identity setting, whereas
Sinkhorn achieves the best EER of 0.66\% and minDCF of 0.085. These results
show that learnable mixing alone is insufficient and that constraining both
the row and column sums is important for effective residual adaptation.

\subsection{Effect of MCRA Placement and Gating}
We next examine the placement and gate design of MCRA using the same
WavLM-Large and ECAPA-TDNN system on VoxCeleb1-O. In the placement
experiment, MCRA uses Sinkhorn normalization, $N=4$, and a learnable gate;
only the residual connections to which MCRA is applied are changed.

\begin{table}[t]
\caption{Effect of MCRA placement on VoxCeleb1-O. Sinkhorn normalization,
a learnable gate, and $N=4$ are fixed for all configurations.}
\label{tab:mcra_placement}
\centering
\setlength{\tabcolsep}{7pt}
\renewcommand{\arraystretch}{1.12}
\begin{tabular}{l|c|c}
\hline
Placement & EER (\%) & minDCF \\
\hline
Self-attention residual only & 0.83 & 0.116 \\
FFN residual only & 0.77 & 0.148 \\
Both attention and FFN & \textbf{0.66} & \textbf{0.085} \\
\hline
\end{tabular}
\end{table}

The self-attention-only setting applies MCRA only to the self-attention
residual connections, whereas the FFN-only setting applies it only to the
FFN residual connections. The third setting applies MCRA to both residual
connections in every Transformer layer. As shown in
Table~\ref{tab:mcra_placement}, using MCRA in both locations achieves the
best EER of 0.66\% and minDCF of 0.085. This result shows that adapting both
residual connections is more effective than adapting either one alone.

We further evaluate the gate that balances the original and mixed residuals.
For this experiment, MCRA is applied to both residual connections with
Sinkhorn normalization and $N=4$.

\begin{table}[t]
\caption{Ablation of the MCRA gate design on VoxCeleb1-O.}
\label{tab:mcra_gate}
\centering
\setlength{\tabcolsep}{7pt}
\renewcommand{\arraystretch}{1.12}
\begin{tabular}{l|c|c|c}
\hline
Gate design & Learnable & EER (\%) & minDCF \\
\hline
No gate & \xmark & 0.77 & 0.105 \\
Fixed gate & \xmark & 0.74 & 0.095 \\
Learnable gate & \cmark & \textbf{0.66} & \textbf{0.085} \\
\hline
\end{tabular}
\end{table}

The no-gate setting removes the interpolation in
Eq.~\eqref{eq:mcra_fusion}, which is equivalent to setting $\lambda_l^c=1$
and using only the mixed residual $\widehat{\vect{r}}_l^c$. The fixed-gate
setting keeps $g_l^c=-6$ throughout training, corresponding to
$\lambda_l^c=\sigma(-6)\approx0.0025$. The learnable-gate setting uses the
same initialization but independently optimizes $g_l^c$ at each residual
connection. Table~\ref{tab:mcra_gate} shows
that the fixed gate improves upon using no gate, and the learnable gate
further reduces the EER to 0.66\% and the minDCF to 0.085. These results
verify the effectiveness of preserving the original residual path and
learning its balance with the adapted path.

\subsection{Visualization of FALA Layer Weights}
\begin{figure}[!t]
    \centering
    \includegraphics[width=0.92\linewidth]{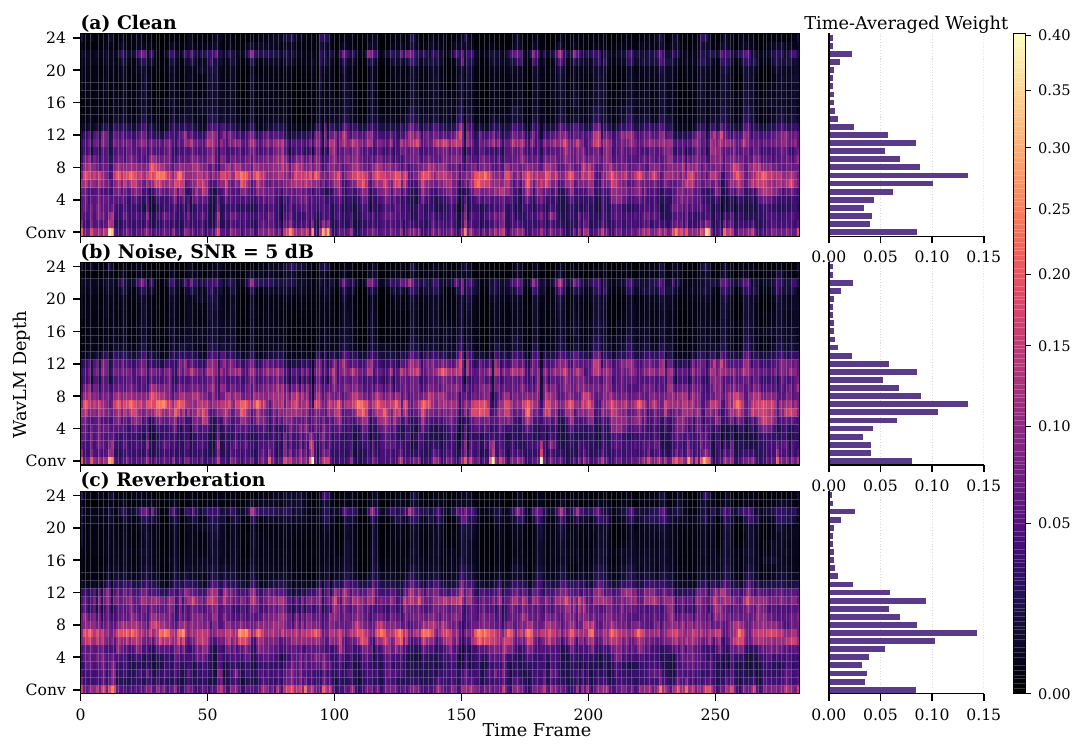}
    \caption{Frame-dependent FALA layer weights under (a) clean speech,
    (b) additive noise at an SNR of 5~dB, and (c) reverberation. The bar plots
    on the right show the layer weights averaged over time.}
    \label{fig:fala_attention_heatmap}
\end{figure}

Fig.~\ref{fig:fala_attention_heatmap} visualizes the layer weights produced
by FALA for the same utterance under clean, noisy, and reverberant
conditions. In each heatmap, every time frame is assigned a different weight
distribution over the WavLM layers, demonstrating the frame-adaptive
property of FALA. The bar plots on the right show the weights averaged over
time. Different layers contribute unequally, and several intermediate layers
receive larger weights. For the same utterance, the overall weight pattern
remains similar after adding noise or reverberation. This result demonstrates
the stability of FALA under acoustic distortions.

\subsection{Visualization}
\begin{figure}[!t]
    \centering
    \includegraphics[width=\linewidth]{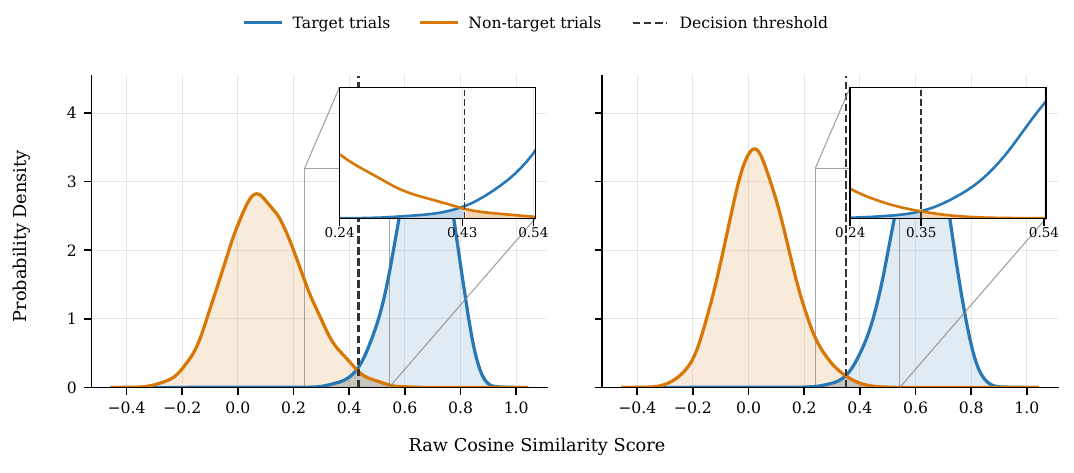}
    \caption{Cosine-score distributions for the (left) frozen WavLM baseline
    and (right) proposed MCRA--FALA system. Blue and orange curves denote
    target and non-target trials, respectively; dashed vertical lines denote
    the decision thresholds. The insets enlarge the overlapping score
    regions.}
    \label{fig:cosine_score}
\end{figure}

Fig.~\ref{fig:cosine_score} compares the cosine-score distributions of
the frozen baseline and MCRA--FALA using the complete VoxCeleb1-O trial
list, which contains 37,720 trials. The blue curve represents target trials, which should receive high scores. The orange curve represents non-target trials, which should receive low scores. The dashed line is the decision threshold. Compared with the baseline,
MCRA--FALA reduces the overlap between the two curves near this threshold.
This makes target and non-target trials easier to distinguish and explains
the lower verification error.

\begin{figure}[!t]
    \centering
    \begin{subfigure}[t]{0.49\linewidth}
        \centering
        \includegraphics[width=\linewidth]{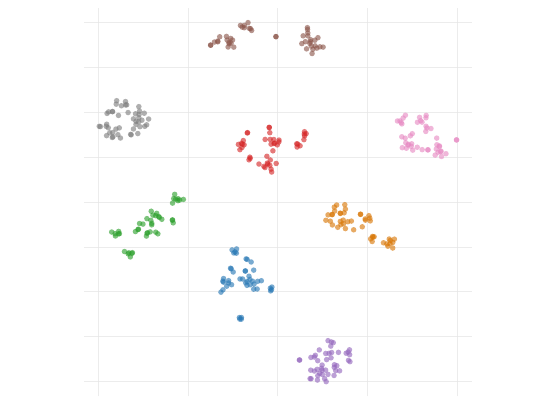}
        \caption{Frozen WavLM baseline.}
    \end{subfigure}
    \hfill
    \begin{subfigure}[t]{0.49\linewidth}
        \centering
        \includegraphics[width=\linewidth]{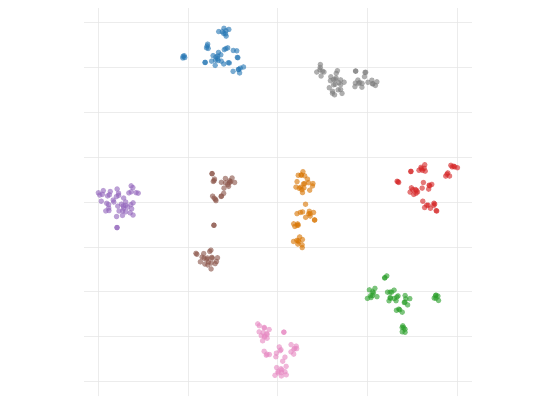}
        \caption{MCRA--FALA system.}
    \end{subfigure}
    \caption{t-SNE visualization of speaker embeddings extracted by
    (a) the frozen WavLM baseline and (b) the proposed MCRA--FALA system.
    The visualization contains eight speakers and 45 utterances per speaker.
    Different colors denote different speakers.}
    \label{fig:embedding_tsne}
\end{figure}

Fig.~\ref{fig:embedding_tsne} shows the corresponding t-SNE visualization.
We select eight speakers and 45 utterances per speaker, giving 360
utterances in total. Embeddings from the same speaker form compact clusters, while different
speakers remain well separated. MCRA--FALA produces clearer cluster
boundaries than the frozen baseline. These visualizations indicate that the
proposed method learns more discriminative speaker representations.

\subsection{Parameter and Computational Efficiency}

\begin{figure}[!t]
    \centering
    \includegraphics[width=\linewidth]{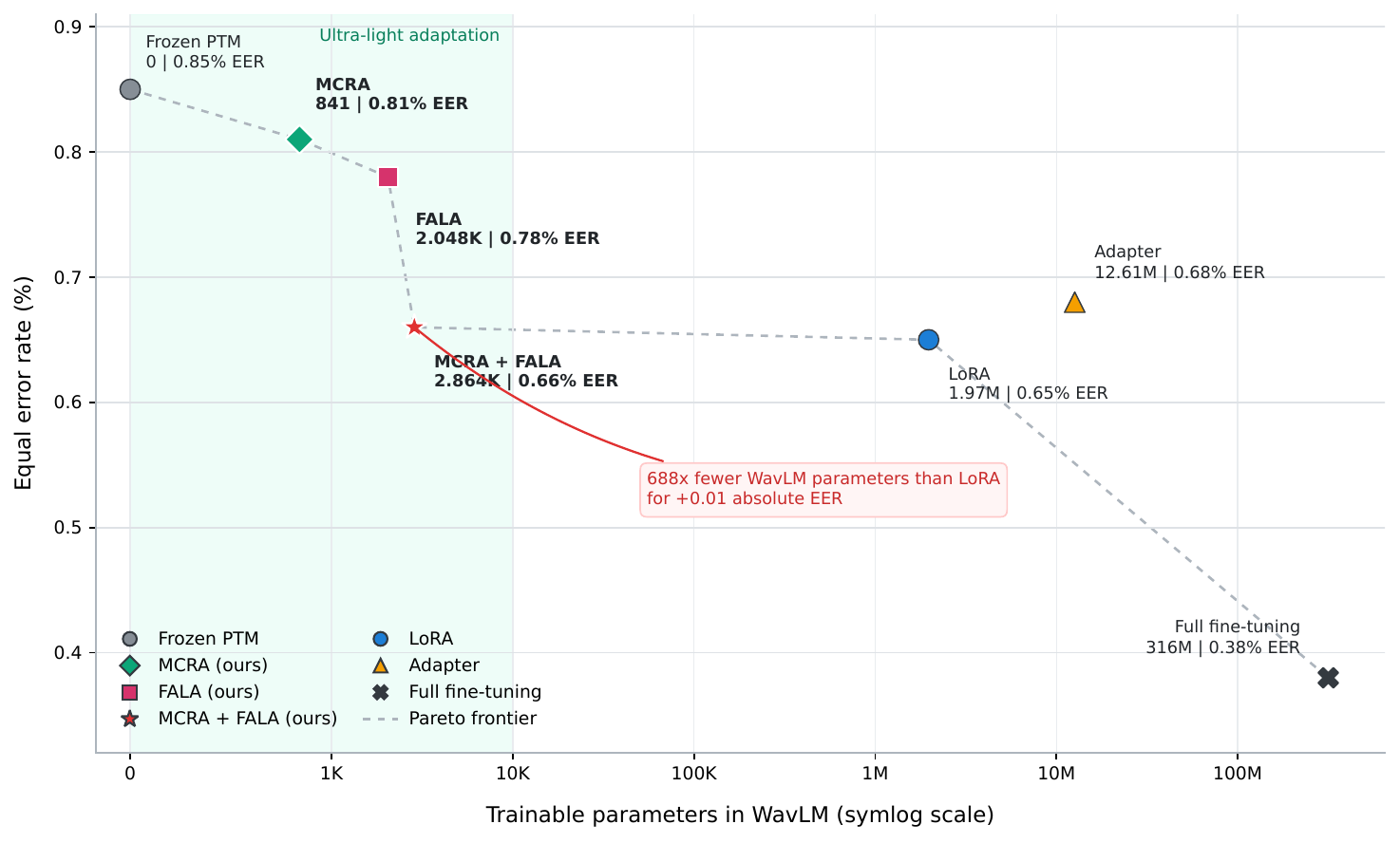}
    \caption{Comparison of trainable parameters in WavLM and EER on
    VoxCeleb1-O. The shaded region indicates methods with fewer than 10K
    trainable parameters.}
    \label{fig:parameter_efficiency}
\end{figure}

Fig.~\ref{fig:parameter_efficiency} compares the number of trainable
parameters in WavLM and the EER of different adaptation methods. A point
closer to the lower-left corner indicates fewer trainable parameters and
lower EER. MCRA and FALA both lie in the ultra-light adaptation region, and
their combination achieves an EER of 0.66\% with only 2.864K trainable
parameters. This result shows that the proposed method provides good speaker
verification performance with very few additional parameters.

\section{Conclusion}
We have proposed the MCRA--FALA, a parameter-efficient framework for
adapting frozen self-supervised speech models to speaker verification. MCRA
partitions each residual state into feature groups and applies
Sinkhorn-normalized mixing to adapt information propagation through the
residual connections. A learnable gate preserves the original residual path
and gradually introduces the adapted path. FALA complements MCRA by
assigning frame-dependent weights to the WavLM layers, allowing each speech
frame to select the most relevant layer representations. Experiments on
VoxCeleb1 and CN-Celeb demonstrate that the two modules are effective
individually and provide complementary improvements when combined. Compared
with the frozen WavLM-Large baseline, MCRA--FALA achieves a relative EER
reduction of 22.4\% on VoxCeleb1-O while introducing only 2.864K trainable
parameters. It also improves performance on CN-Celeb with only 5.168K
parameters. Ablation experiments confirm the effectiveness of the Sinkhorn
constraint, applying MCRA to both residual connections, and using a
learnable gate. The visualization further shows that FALA performs
frame-adaptive layer selection and maintains stable layer-weight patterns
under noise and reverberation. These results demonstrate that adapting
residual information propagation and layer aggregation provides an
effective and highly parameter-efficient approach to speaker verification.



\balance

\begin{IEEEbiography}[{\includegraphics[width=1in,height=1.25in,clip,keepaspectratio]{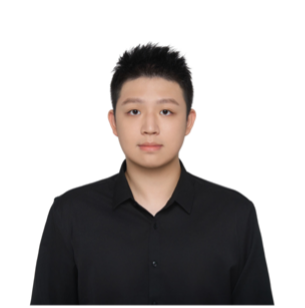}}]{Zezhong Jin} (Student Member, IEEE) received the B.Eng. degree in electronic information engineering from Hebei University, Baoding, China, in 2021, and the M.Sc. degree in electronic and information engineering from The Hong Kong Polytechnic University, Hong Kong, in 2023. He received the Ph.D. degree from The Hong Kong Polytechnic University in 2026. His research focuses on speaker verification, self-supervised learning, knowledge distillation, and audio large language models. He interned at Microsoft Research Asia (MSRA) in 2025 and at ByteDance in 2026 through its Talent Internship Program.\end{IEEEbiography}
\begin{IEEEbiography}[{\includegraphics[width=1in,height=1.25in,clip,keepaspectratio]{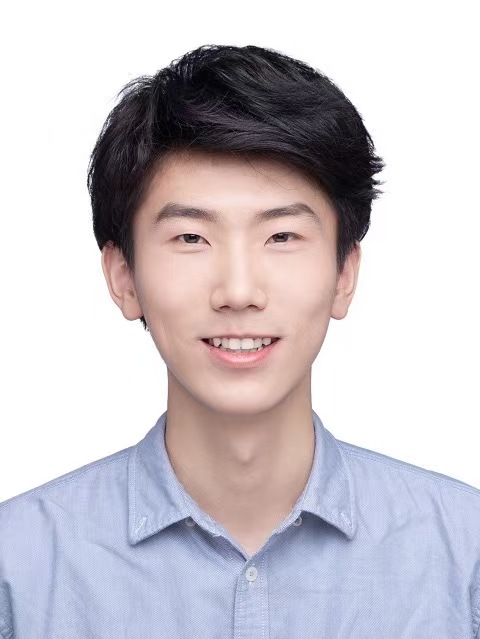}}]{Xiaoyu Wang} received the B.Eng. degree in information engineering from the Chien-Shiung Wu College, Southeast University, Nanjing, China, in 2021, and the M.Eng. degree in electronic and information engineering from Tsinghua Shenzhen International Graduate School, Tsinghua University, Shenzhen, China, in 2024. He is currently a Senior Algorithm Engineer at Baidu, working on large language models. His research interests include reinforcement learning for large language models, multimodal large language models, speaker verification, and knowledge distillation.\end{IEEEbiography}
\begin{IEEEbiography}[{\includegraphics[width=1in,height=1.25in,clip,keepaspectratio]{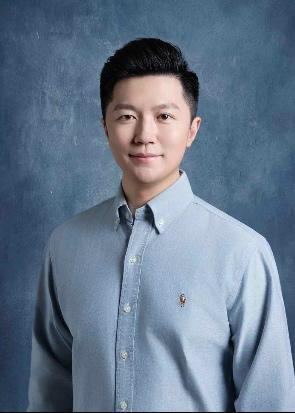}}]{Zhe LI} is currently a postdoctoral researcher with the Department of Electrical Engineering, Stanford University. He previously served as a postdoctoral fellow at The University of Hong Kong (HKU). His research focuses on speech large language models (Speech LLMs) and robust speaker representation learning, with broader interests in multimodal AI for healthcare applications. He received the Ph.D. degree from the Department of Electrical and Electronic Engineering at The Hong Kong Polytechnic University (PolyU) in 2025. He received the M.Sc. degree in Software Engineering from Xinjiang University, Urumqi, China, in 2021, and the B.Eng. degree in Computer Science from Qilu University of Technology, Jinan, China, in 2016. He was a research intern at Microsoft Research Asia (MSRA) in 2025. He previously conducted international collaborative research as a visiting student scholar with the Department of Electrical Engineering, Stanford University, in 2024. He received the 2020 Excellent Science and Technology Achievement Award from the Chinese Association for Artificial Intelligence, and his co-authored paper received the Best Student Paper Runner-Up Award at PRICAI 2024.\end{IEEEbiography}
\begin{IEEEbiography}[{\includegraphics[width=1in,height=1.25in,clip,keepaspectratio]{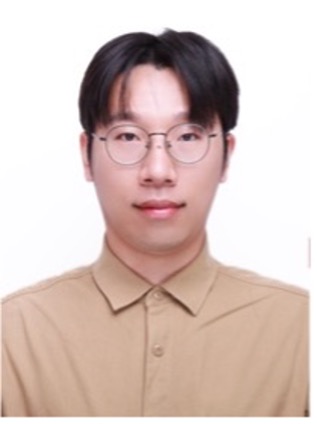}}]{Chong-Xin Gan} received his B.Eng. degree in Computer Science and Technology from Zhengzhou University in 2019 and M.Sc. degree in Electronic and Information Engineering from The Hong Kong Polytechnic University in 2021. He is currently pursuing a Ph.D. degree in the Department of Electrical and Electronic Engineering at The Hong Kong Polytechnic University. He was also a visiting student at The Center for Speech Technology Research, The University of Edinburgh. His research interests include speaker verification, sound event classification, and deep learning.\end{IEEEbiography}
\begin{IEEEbiography}[{\includegraphics[width=1in,height=1.25in,clip,keepaspectratio]{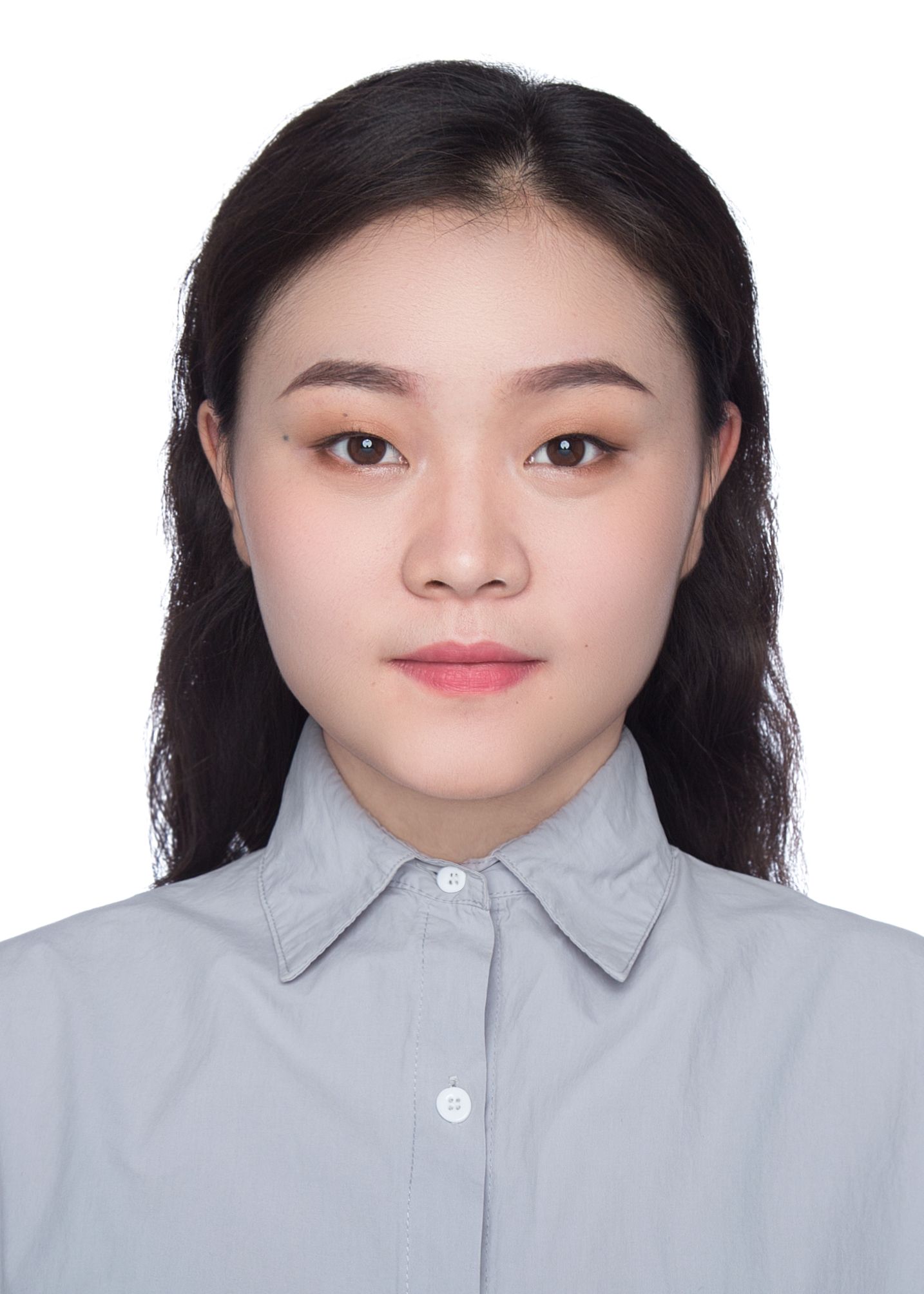}}]{QIN Siqing} is currently a Ph.D. student with the Department of Electrical and Electronic Engineering, The Hong Kong Polytechnic University. Her research focuses on speech deepfake detection and speaker embeddings. She received the M.Sc. degree in computer science and technology from Tianjin University, Tianjin, China, in 2022, and the B.Eng. degree in computer science from Zhengzhou University, Henan, China, in 2015. She is conducting international collaborative research as a visiting student scholar with the Department of Electrical and Computer Engineering, Johns Hopkins University.\end{IEEEbiography}
\begin{IEEEbiography}[{\includegraphics[width=1in,height=1.25in,clip,keepaspectratio]{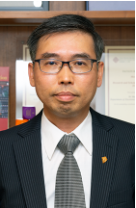}}]{Man-Wai Mak} (Senior Member, IEEE) received the B.Eng. (with Hons.) degree in electronic engineering from Newcastle Upon Tyne Polytechnic, Newcastle upon Tyne, U.K., in 1989, and the Ph.D. degree in electronic engineering from Northumbria University, Newcastle upon Tyne (formerly University of Northumbria), in 1993. In 1993, he joined the Department of Electronic and Information Engineering (EIE), The Hong Kong Polytechnic University, Hong Kong. From 2021 to 2023, he was the Interim Head with EIE. He is currently a Professor and Associate Head with the Department of Electronic and Electrical Engineering. He has authored more than 220 technical articles and books in speaker recognition, machine learning, bioinformatics, and biomedical engineering. He is the coauthor of the postgraduate textbook \emph{Biometric Authentication: A Machine Learning Approach} (Prentice-Hall, 2005), \emph{Machine Learning for Protein Subcellular Localization Prediction} (De Gruyter, 2015), and \emph{Machine Learning for Speaker Recognition} (Cambridge University Press, 2020). His research interests include speaker recognition, machine learning, spoken language processing, biomedical engineering, and bioinformatics. He was a Guest Editor of international journals and an Associate Editor for IEEE TRANSACTIONS ON AUDIO, SPEECH, AND LANGUAGE PROCESSING, Journal of Signal Processing Systems, Advances in Artificial Neural Systems, and IEEE BIOMETRICS COMPENDIUM. He was the Tutorial Speaker at Interspeech 2016. He received three Faculty of Engineering Research Grant Achievement Awards and a Faculty Award for Outstanding Performance in Research and Scholarly Activities. He was an Executive Committee Member of the IEEE Hong Kong Section Computer Chapter from 1995 to 2007 and Chairperson from 2003 to 2005. He served on the IEEE Machine Learning for Signal Processing Technical Committee from 2005 to 2007 and was a Technical Committee Member of the IEEE Computational Intelligence Society, Intelligent Systems Applications, in 2008. He was Area Chair of Interspeech 2014 and ICTAI 2016, a Steering Committee Member of ISCSLP and ISCSLP 2016, and Program Co-Chair of ISCSLP 2018 and ISCSLP 2021.\end{IEEEbiography}
\begin{IEEEbiography}[{\includegraphics[width=1in,height=1.25in,clip,keepaspectratio]{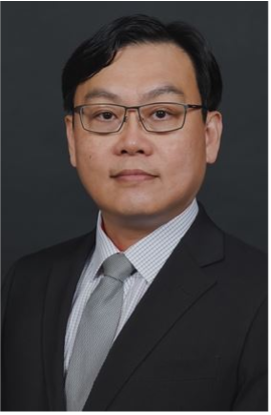}}]{Kong Aik Lee} (Senior Member, IEEE) received the Ph.D. degree from Nanyang Technological University, Singapore, in 2006. From 2006 to 2018, he was a Research Scientist and then a Strategic Planning Manager (concurrent appointment) with the Institute for Infocomm Research, Singapore. From 2018 to 2020, he was a Senior Principal Researcher with the Data Science Research Laboratories, NEC Corporation, Tokyo, Japan. He was an Associate Professor with the Singapore Institute of Technology, Singapore, while holding a concurrent appointment as a Principal Scientist and a Group Leader with the Agency for Science, Technology and Research (A*STAR), Singapore. He is currently an Associate Professor with The Hong Kong Polytechnic University, Hong Kong. His research interests include the automatic and paralinguistic analysis of speaker characteristics, ranging from speaker recognition, language and accent recognition, voice biometrics, spoofing, and countermeasures. From 2017 to 2021, he was an Associate Editor for IEEE/ACM TRANSACTIONS ON AUDIO, SPEECH, AND LANGUAGE PROCESSING. Since 2016, he has been an Editorial Board Member of Computer Speech and Language. He is an elected Member of the IEEE Speech and Language Processing Technical Committee and was the General Chair of the Speaker Odyssey 2020 Workshop. He received the Singapore IES Prestigious Engineering Achievement Award in 2013 and the Outstanding Service Award from IEEE ICME in 2020.\end{IEEEbiography}

\end{document}